\documentclass{article}
\usepackage[margin=1.9cm]{geometry}
\usepackage{amsmath,amssymb,amsthm,amsfonts}
\usepackage[utf8]{inputenc}
\usepackage{graphicx}
\usepackage{enumitem}
\usepackage[justification=centering]{caption}
\usepackage{url}
\usepackage{multicol}
\usepackage[noline,noend]{algorithm2e}
    \SetArgSty{textnormal}
    \SetNoFillComment
    \SetCommentSty{Xcmfont}

\setlist{nosep}
\setlistdepth{9}
\setlist[enumerate,1]{label=$\arabic*.$}
\setlist[enumerate,2]{label=$\alph*.$}
\setlist[enumerate,3]{label=$\roman*.$}
\setlist[enumerate,4]{label=$(\arabic*)$}
\setlist[enumerate,5]{label=$(\alph*)$}
\setlist[enumerate,6]{label=$(\roman*)$}
\setlist[enumerate,7]{label=$\arabic*$}
\setlist[enumerate,8]{label=$\alph*$}
\setlist[enumerate,9]{label=$\roman*$}
\renewlist{enumerate}{enumerate}{9}

\newenvironment{Xfig}
    {\par\medskip\noindent\minipage{\linewidth}\begin{center}}
    {\end{center}\endminipage\par\medskip}
\newenvironment{Xtab}
    {\par\medskip\noindent\minipage{\linewidth}\begin{center}}
    {\end{center}\endminipage\par\medskip}

\newcommand{\Xset}{\!\leftarrow\!}
\newcommand{\Xund}{\rule{.4em}{.4pt}} 
\newcommand{\Xin}{\!\in\!}
\newcommand{\Xeq}{\!=\!}
\newcommand{\Xlb}{[\![}
\newcommand{\Xrb}{]\!]}

\newcommand{\XB}{\mathcal{B}}

\newcommand{\XF}{\mathcal{F}}
\newcommand{\XI}{\mathcal{I}}
\newcommand{\XL}{\mathcal{L}}
\newcommand{\XN}{\mathcal{N}}

\newcommand{\XS}{\mathcal{S}}
\newcommand{\XT}{\mathcal{T}}
\newcommand{\XX}{\mathcal{X}}
\newcommand{\YB}{\mathbb{B}}
\newcommand{\YF}{\mathbb{F}}
\newcommand{\YN}{\mathbb{N}}

\newcommand{\YQ}{\mathbb{Q}}

\newcommand{\Xstirling}[2]{\genfrac{\{}{\}}{0pt}{}{#1}{#2}}

\makeatletter
\newcommand*{\Relbarfill@}{\arrowfill@\Relbar\Relbar\Relbar}
\newcommand*{\Xlongeq}[2][]{\ext@arrow 0055\Relbarfill@{#1}{\text{#2}}}
\newcommand*{\Xbar}[1]{\overline{#1\vphantom{\bar{#1}}}}
\makeatother

\theoremstyle{definition}
\newtheorem{Xdef}{Definition}
\newtheorem{XThe}{Theorem}
\newtheorem{XLem}{Lemma}
\newtheorem{Xobs}{Observation}

\title{Tagged Deterministic Finite Automata with Lookahead}
\author{Ulya Trofimovich\\
\texttt{\small{skvadrik@gmail.com}}}
\date{August 2017}

\begin{document}

\maketitle

\begin{abstract}
\noindent
This paper extends the work of Laurikari \cite{Lau00} \cite{Lau01}
and Kuklewicz \cite{Kuk07} on tagged deterministic finite automata (TDFA)
in the context of submatch extraction in regular expressions.
The main goal of this work is application of TDFA to lexer generators that optimize for speed of the generated code.
I suggest a number of practical improvements to Laurikari algorithm;
notably, the use of one-symbol lookahead, which results in significant reduction of tag variables and operations on them.
Experimental results confirm that lookahead-aware TDFA are considerably faster and usually smaller than baseline TDFA;
and they are reasonably close in speed and size to ordinary DFA used for recognition of regular languages.
The proposed algorithm can handle repeated submatch and therefore is applicable to full parsing.
Furthermore, I examine the problem of disambiguation in the case of leftmost greedy and POSIX policies.
I formalize POSIX disambiguation algorithm suggested by Kuklewicz
and show that the resulting TDFA are as efficient as Laurikari TDFA or TDFA that use leftmost greedy disambiguation.
All discussed algorithms are implemented in the open source lexer generator RE2C.
\end{abstract}
\vspace{1em}

\begin{multicols}{2}

\section*{Introduction}

RE2C is a lexer generator for C: it compiles regular expressions into C code \cite{BC93} \cite{RE2C}.
Unlike regular expression libraries, lexer generators separate compilation and execution steps:
they can spend considerable amount of time on compilation in order to optimize the generated code.
Consequently, lexer generators are usually aimed at generating efficient code rather than supporting multiple extensions;
they use deterministic automata and avoid features that need more complex computational models.
In particular, RE2C aims at generating lexers that are at least as fast as reasonably optimized hand-coded lexers.
It compiles regular expressions into deterministic automata,
applies a number of optimizations to reduce automata size
and converts them directly into C code in the form of conditional jumps:
this approach results in more efficient and human-readable code than table-based lexers.
In addition, RE2C has a flexible interface:
instead of using a fixed program template,
it lets the programmer define most of the interface code
and adapt the lexer to a particular environment.
\\ \\
One useful extension of traditional regular expressions that cannot be implemented using ordinary DFA is submatch extraction and parsing.
Many authors studied this subject and developed algorithms suitable for their particular settings and problem domains.
Their approaches differ in various respects:
the specific subtype of problem (full parsing, submatch extraction with or without history of repetitions),
the underlying formalism (backtracking,
nondeterministic automata, deterministic automata, 
multiple automata, lazy determinization),
the number of passes over the input (streaming, multi-pass),
space consumption with respect to input length (constant, linear),
handing of ambiguity (unhandled, manual disambiguation, default disambiguation policy, all possible parse trees), etc.
Most of the algorithms are unsuitable for RE2C: they are either insufficiently generic (cannot handle ambiguity),
or too heavyweight (incur overhead on regular expressions with only a few submatches or no submatches at all).
Laurikari algorithm is outstanding in this regard.
It is based on a single deterministic automaton, runs in one pass and requires linear time,
and the consumed space does not depend on the input length.
What is most important, the overhead on submatch extraction depends on the detalization of submatch:
on submatch-free regular expressions Laurikari automaton reduces to a simple DFA.
\\ \\
From RE2C point of view this is close enough to hand-written code:
you only pay for what you need, like a reasonable programmer would do.
However, a closer look at Laurikari automata reveals that
they behave like a very strange programmer who is unable to think even one step ahead.
Take, for example, regular expression \texttt{a*b*}
and suppose that we must find the position between \texttt{a} and \texttt{b} in the input string.
The programmer would probably match all \texttt{a}, then save the input position, then match all \texttt{b}:

\begin{Xfig}
\begin{small}
\begin{verbatim}
    while (*s++ == 'a') ;
    p = s;
    while (*s++ == 'b') ;
\end{verbatim}
\end{small}
\end{Xfig}

And this corresponds to automaton behavior:

\begin{Xfig}
\begin{small}
\begin{verbatim}
    p = s;
    while (*s++ == 'a') p = s;
    while (*s++ == 'b') ;
\end{verbatim}
\end{small}
\end{Xfig}

This behavior is correct (it yields the same result), but strangely inefficient:
it repeatedly saves input position after every \texttt{a},
while for the programmer it is obvious that there is nothing to save until the first non-\texttt{a}.
One might object that the C compiler would optimize out the difference,
and it probably would in simple cases like this.
However, the flaw is common to all Laurikari automata:
they ignore lookahead when recording submatches.
But they don't have to; with a minor fix we can teach them
to delay recording until the right lookahead symbol shows up.
This minor fix is my first contribution.
\\ \\
Another problem that needs attention is disambiguation.
The original paper \cite{Lau01} claims to have POSIX semantics, but it was proved to be wrong \cite{LTU}.
Since then Kuklewicz suggested a fix for Laurikari algorithm that does have POSIX semantics \cite{Regex-TDFA}, but he never formalized the resulting algorithm.
The informal description \cite{Kuk07} is somewhat misleading as it suggests that Kuklewicz automata
require additional run-time operations to keep track of submatch history and hence are less efficient than Laurikari automata.
That is not true, as we shall see: all the added complexity is related to determinization,
while the resulting automata are just the same (except they have POSIX semantics).
Kuklewicz did not emphasize this, probably because his implementation constructs TDFA lazily at run-time.
I formalize Kuklewicz algorithm; this is my second contribution.
\\ \\
Finally, theory is no good without practice.
Even lookahead-aware automata contain redundant operations
which can be reduced by basic optimizations like liveness analysis and dead code elimination.
The overall number of submatch records can be minimized using technique similar to register allocation.
I suggest another tweak of Laurikari algorithm that makes optimizations particularly easy
and show that they are useful even in the presence of an optimizing C compiler.
RE2C implementation of submatch extraction is the motivation and the main goal of this work.
\\ \\
The rest of this paper is arranged as follows.
We start with theoretical foundations and gradually move towards practical algorithms.
Section \ref{section_regular_expressions} revises the basic definition of regular expressions.
In section \ref{section_tagged_extension} we extend it with tags
and define ambiguity with respect to submatch extraction.
In section \ref{section_tnfa} we convert regular expressions to nondeterministic automata
and in section \ref{section_closure} study various algorithms for $\epsilon$-closure construction.
Section \ref{section_disambiguation} tackles disambiguation problem;
we discuss leftmost greedy and POSIX policies and the necessary properties that disambiguation policy should have in order to allow efficient submatch extraction.
Section \ref{section_determinization} is the main part of this paper: it describes determinization algorithm.
Section \ref{section_implementation} highlights some practical implementation details and optimizations.
Section \ref{section_tests_and_benchmarks} concerns correctness testing and benchmarks.
Finally, section \ref{section_conclusions} contains conclusions
and section \ref{section_future_work} points directions for future work.

\section{Regular expressions}\label{section_regular_expressions}

Regular expressions are a \emph{notation} that originates in the work of Kleene
\emph{``Representation of Events in Nerve Nets and Finite Automata''} \cite{Kle51} \cite{Kle56}.
He used this notation to describe \emph{regular events}:
each regular event is a set of \emph{definite events},
and the class of all regular events is defined inductively
as the least class containing basic events (empty set and all unit sets)
and closed under the operations of \emph{sum}, \emph{product} and \emph{iterate}.
Kleene showed that regular events form exactly the class of events that can be represented by McCulloch-Pitts nerve nets or, equivalently, finite automata.
However, generalization of regular events to other fields of mathematics remained an open problem;
in particular, Kleene raised the question whether regular events could be reformulated
as a deductive system based on logical axioms and algebraic laws.
This question was thoroughly investigated by many authors (see \cite{Koz94} for a historic overview)
and the formalism became known as \emph{the algebra of regular events} 
or, more generally, the \emph{Kleene algebra} $\mathcal{K} \Xeq (K, +, \cdot, *, 1, 0)$.
Several different axiomatizations of Kleene algebra were given;
in particular, Kozen gave a finitary axiomatization based on equations and equational implications and sound for all interpretations \cite{Koz94}.
See also \cite{Gra15} for extensions of Kleene algebra and generalization to the field of context-free languages.
\\

The following definition of regular expressions, with minor notational differences, is widely used in literature
(see e.g. [HU90], page 28, or [SS88], page 67):

    \begin{Xdef}\label{re}
    \emph{Regular expression (RE)} over finite alphabet $\Sigma$ is one of the following:
    \begin{enumerate}
        \medskip
        \item[] $\emptyset$, $\epsilon$ and $\alpha \Xin \Sigma$ (\emph{atomic} RE)
        \item[] $(e_1 | e_2)$, where $e_1$, $e_2$ are RE over $\Sigma$ (\emph{sum})
        \item[] $(e_1 e_2)$,   where $e_1$, $e_2$ are RE over $\Sigma$ (\emph{product})
        \item[] $(e^*)$,       where $e$ is a RE over $\Sigma$ (\emph{iteration})
        \medskip
    \end{enumerate}
    \end{Xdef}

The usual assumption is that iteration has precedence over product and product has precedence over sum,
and redundant parentheses may be omitted.
$\emptyset$ and $\epsilon$ are special symbols not included in the alphabet $\Sigma$ (they correspond to $1$ and $0$ in the Kleene algebra).
Since RE are only a notation, their exact meaning depends on the particular \emph{interpretation}.
In the \emph{standard} interpretation RE denote \emph{languages}: sets of strings over the alphabet of RE.
\\

Let $\epsilon$ denote the \emph{empty string} (not to be confused with RE $\epsilon$),
and let $\Sigma^*$ denote the set of all strings over $\Sigma$ (including the empty string $\epsilon$).

    \begin{Xdef}
    \emph{Language} over $\Sigma$ is a subset of $\Sigma^*$.
    \end{Xdef}

    \begin{Xdef}\label{langunion}
    \emph{Union} of two languages $L_1$ and $L_2$ is
    $L_1 \cup L_2 = \{ x \mid x \Xin L_1 \vee x \Xin L_2 \}$
    \end{Xdef}

    \begin{Xdef}\label{langproduct}
    \emph{Product} of two languages $L_1$ and $L_2$ is
    $L_1 \cdot L_2 = \{ x_1 x_2 \mid x_1 \Xin L_1 \wedge x_2 \Xin L_2 \}$
    \end{Xdef}

    \begin{Xdef}\label{langiterate}
    n-Th \emph{Power} of language $L$ is
    $$L^n = \begin{cases}
            \{ \epsilon \} & \text{if } n \Xeq 0 \\[-0.5em]
            L \cdot L^{n - 1} & \text{if } n\!>\!0
        \end{cases}$$
    \end{Xdef}

    \begin{Xdef}\label{langiterate}
    \emph{Iterate} of language $L$ is
    $L^* = \bigcup\limits_{n = 0}^\infty L^n$.
    \end{Xdef}

    \begin{Xdef}
    \emph{(Language interpretation of RE)} \\
    RE denotes a language over $\Sigma$:
    \begin{align*}
        \XL \Xlb \emptyset \Xrb &= \emptyset \\
        \XL \Xlb \epsilon \Xrb &= \{ \epsilon \} \\
        \XL \Xlb \alpha \Xrb &= \{\alpha\} \\
        \XL \Xlb e_1 | e_2 \Xrb &= \XL \Xlb e_1 \Xrb \cup \XL \Xlb e_2 \Xrb \\
        \XL \Xlb e_1 e_2 \Xrb &= \XL \Xlb e_1 \Xrb \cdot \XL \Xlb e_2 \Xrb \\
        \XL \Xlb e^* \Xrb &= \XL \Xlb e \Xrb ^*
    \end{align*}
    \end{Xdef}

Other interpretations are also possible;
one notable example is the \emph{type interpretation},
in which RE denote sets of parse trees \cite{BT10} \cite{Gra15}.
This is close to what we need for submatch extraction,
except that we are interested in partial parse structure rather than full parse trees.

    \begin{Xdef}
    Language $L$ over $\Sigma$ is \emph{regular} iff exists RE $e$ over $\Sigma$
    such that $L$ is denoted by $e$: $\XL \Xlb e \Xrb \Xeq L$.
    \end{Xdef}

For the most useful RE there are special shortcuts:
    \begin{align*}
        e^n     &\quad\text{for}\quad \overbrace{e \dots e}^{n} \\[-0.5em]
        e^{n,m} &\quad\text{for}\quad e^n | e^{n+1} | \dots | e^{m-1} | e^m \\[-0.5em]
        e^{n,}  &\quad\text{for}\quad e^n e^* \\[-0.5em]
        e^+     &\quad\text{for}\quad ee^* \\[-0.5em]
        e^?     &\quad\text{for}\quad e | \epsilon
    \end{align*}

\section{Tagged extension}\label{section_tagged_extension}

In short, tags are position markers attached to the structure of RE.
The idea of adding such markers is not new:
many RE flavors have \emph{capturing groups}, or the \emph{lookahead} operator, or \emph{pre-} and \emph{post-context} operators;
all of them are used for submatch extraction of some sort.\footnote{
Position markers in RE are sometimes used in a different sence:
Watson mentions the \emph{dotted} RE \cite{Wat93}
that go back to DeRemers's construction of DFA, which originates in LR parsing invented by Knuth.
The \emph{dot} itself is the well-known LR \emph{item} which separates the already parsed and yet unparsed parts of the rule.
}
Laurikari used the word \emph{tag}.
He did not define tags explicitly; rather, he defined automata with tagged transitions.
We take a slightly different approach, inspired by \cite{BT10}, \cite{Gra15} and a number of other publications.
First, we define an extension of RE: tagged RE,
and two interpretations: \emph{S-language} that ignores tags and \emph{T-language} that preserves them.
T-language has the bare minimum of information necessary for submatch extraction;
in particular, it is less expressive than \emph{parse trees} or \emph{types} that are used for RE parsing.
Then we define \emph{ambiguity} and \emph{disambiguation policy} in terms of relations between the two interpretations.
Finally, we show how T-language can be converted to \emph{tag value functions} used by Laurikari
and argue that the latter representation is insufficient as it cannot express ambiguity in certain RE.
\\

In tagged RE we use generalized repetition $e^{n,m}$ (where $0 \!\leq\! n \!\leq\! m \!\leq\! \infty$)
instead of iteration $e^*$ as one of the three base operations.
The reason for this is the following:
bounded repetition cannot be expressed in terms of union and product without duplication of RE,
and duplication may change semantics of submatch extraction.
For example, POSIX RE \texttt{(a(b?))\{2\}} contains two submatch groups (aside from the whole RE),
but if we rewrite it as \texttt{(a(b?))(a(b?))}, the number of submatch groups will change to four.
Generalized repetition, on the other hand, allows to express all kinds of iteration without duplication.
\\

    \begin{Xdef}\label{tre}
    \emph{Tagged regular expression (TRE)} over disjoint finite alphabets $\Sigma$ and $T$ is one of the following:
    \begin{enumerate}
        \medskip
        \item[] $\emptyset$, $\epsilon$, $\alpha \Xin \Sigma$ and $t \Xin T$ (\emph{atomic} TRE)
        \item[] $(e_1 | e_2)$, where $e_1$, $e_2$ are TRE over $\Sigma$, $T$ (\emph{sum})
        \item[] $(e_1 e_2)$,   where $e_1$, $e_2$ are TRE over $\Sigma$, $T$ (\emph{product})
        \item[] $(e^{n,m})$,   where $e$ is a TRE over $\Sigma$, $T$ \\
            \hphantom{\qquad} and $0 \!\leq\! n \!\leq\! m \!\leq\! \infty$ (\emph{repetition})
        \medskip
    \end{enumerate}
    \end{Xdef}

    As usual, we assume that repetition has precedence over product and product has precedence over sum,
    and redundant parentheses may be omitted.
    Additionally, the following shorthand notation may be used:
    \begin{align*}
        e^*     &\quad\text{for}\quad e^{0,\infty} \\[-0.5em]
        e^+     &\quad\text{for}\quad e^{1,\infty} \\[-0.5em]
        e^?     &\quad\text{for}\quad e^{0,1} \\[-0.5em]
        e^n     &\quad\text{for}\quad e^{n,n}
    \end{align*}

    \begin{Xdef}
    TRE over $\Sigma$, $T$ is \emph{well-formed} iff
    all tags in it are pairwise different
    and $T \Xeq \{ 1, \dots, |T| \}$.
    We will consider only well-formed TRE.
    \end{Xdef}

If we assume that tags are aliases to $\epsilon$, then every TRE over $\Sigma$, $T$ is a RE over $\Sigma$:
intuitively, this corresponds to erasing all submatch information. 
We call this \emph{S-language interpretation} (short from ``sigma'' or ``source''),
and the corresponding strings are \emph{S-strings}:

    \begin{Xdef}\label{defslang}
    \emph{(S-language interpretation of TRE)} \\
    TRE over $\Sigma$, $T$ denotes a language over $\Sigma$:
    \begin{align*}
        \XS \Xlb \emptyset \Xrb &= \emptyset \\
        \XS \Xlb \epsilon \Xrb &= \{ \epsilon \} \\
        \XS \Xlb \alpha \Xrb &= \{\alpha\} \\
        \XS \Xlb t \Xrb &= \{\epsilon\} \\
        \XS \Xlb e_1 | e_2 \Xrb &= \XS \Xlb e_1 \Xrb \cup \XS \Xlb e_2 \Xrb \\
        \XS \Xlb e_1 e_2 \Xrb &= \XS \Xlb e_1 \Xrb \cdot \XS \Xlb e_2 \Xrb \\
        \XS \Xlb e^{n,m} \Xrb &= \bigcup\limits_{i=n}^m \XS \Xlb e \Xrb ^i
    \end{align*}
    \end{Xdef}

On the other hand, if we interpret tags as symbols, then every TRE over $\Sigma$, $T$ is a RE over the joined alphabet $\Sigma \cup T$.
This interpretation retains submatch information; however, it misses one important detail: \emph{negative} submatches.
Negative submatches are implicitly encoded in the structure of TRE:
we can always deduce the \emph{absence} of tag from its \emph{presence} on alternative branch of TRE.
To see why this is important, consider POSIX RE \texttt{(a(b)?)*} matched against string \texttt{aba}.
The outermost capturing group matches twice at offsets 0, 2 and 2, 3 (opening and closing parentheses respectively).
The innermost group matches only once at offsets 1, 2; there is no match corresponding to the second outermost iteration.
POSIX standard demands that the value on the last iteration is reported: that is, the absence of match.
Even aside from POSIX, one might be interested in the whole history of submatch.
Therefore we will rewrite TRE in a form that makes negative submatches explicit
(by tracking tags on alternative branches and inserting negative tags at all join points).
Negative tags are marked with bar, and $\Xbar{T}$ denotes the set of all negative tags.

    \begin{Xdef}\label{deftlang}
    Operator $\XX$ rewrites TRE over $\Sigma$, $T$ to a TRE over $\Sigma$, $T \cup \Xbar{T}$:
    \begin{align*}
        \XX(\emptyset) &= \emptyset \\
        \XX(\epsilon) &= \epsilon \\
        \XX(\alpha) &= \alpha \\
        \XX(t) &= t \\
        \XX(e_1 | e_2)
            &=      \XX(e_1) \chi(e_2) \mid \XX(e_2) \chi(e_1) \\
        \XX(e_1 e_2) &= \XX(e_1) \XX(e_2) \\
        \XX(e^{n,m}) &= \begin{cases}
                \XX(e)^{1,m} \mid \chi(e) &\text{if } n \Xeq 0 \\
                \XX(e)^{n,m} &\text{if } n \!\geq\! 1
            \end{cases} \\
        \\
        \text{where }
            \chi(e) &= \Xbar{t_1} \dots \Xbar{t_n}, \text{ such that} \\
                &t_1 \dots t_n \text{ are all tags in } e
    \end{align*}
    \end{Xdef}

    \begin{Xdef}\label{deftlang}
    \emph{(T-language interpretation of TRE)} \\
    TRE over $\Sigma$, $T$ denotes a language over $\Sigma \cup T \cup \Xbar{T}$:
    $\XT \Xlb e \Xrb = \XL \Xlb \widetilde{e} \Xrb$, where $\widetilde{e}$ is a RE
    syntactically identical to TRE $\XX(e)$.
    \end{Xdef}

The language over $\Sigma \cup T \cup \Xbar{T}$ is called \emph{T-language}
(short from ``tag'' or ``target''), and its strings are called \emph{T-strings}.
For example:
\begin{align*}
    \XT \Xlb \beta &| (\alpha 1)^{0,2} \Xrb
=
    \XT \Xlb \beta \Xbar{1} | \big( (\alpha 1)^{1,2} | \Xbar{1} \big) \Xrb = \\
&=
    \XT \Xlb \beta \Xbar{1} \Xrb
        \cup \XT \Xlb (\alpha 1)^{1,2} \Xrb
        \cup \XT \Xlb \Xbar{1} \Xrb = \\
&=
    \XT \Xlb \beta \Xrb
        \cdot \XT \Xlb \Xbar{1} \Xrb \cup
    \XT \Xlb \alpha 1 \Xrb
            \cup \XT \Xlb \alpha 1 \Xrb \cdot \XT \Xlb \alpha 1 \Xrb
        \cup \{\Xbar{1}\} = \\
&=
    \{\beta \Xbar{1}\} \cup
    \{\alpha 1 \} \cup
    \{\alpha 1 \alpha 1 \} \cup
    \{\Xbar{1}\} = \\
&=
    \{\beta \Xbar{1}, \Xbar{1}, \alpha 1, \alpha 1 \alpha 1 \}
\end{align*}

    \begin{Xdef}\label{untag}
    The \emph{untag} function $S$ converts T-strings into S-strings:
    $S(\gamma_0 \dots \gamma_n) \Xeq \alpha_0 \dots \alpha_n$, where:
    $$\alpha_i = \begin{cases}
            \gamma_i &\text{if } \gamma_i \Xin \Sigma \\[-0.5em]
            \epsilon &\text{otherwise}
        \end{cases}$$
    \end{Xdef}

It is easy to see that for any TRE $e$, $\XS \Xlb e \Xrb$
is exactly the same language as $\{S(x) \mid x \Xin \XT \Xlb e \Xrb\}$.
Moreover, the relation between S-language and T-language
describes exactly the problem of submatch extraction:
given a TRE $e$ and an S-string $s \Xin \XS \Xlb e \Xrb$,
find the corresponding T-string $x \Xin \XT \Xlb e \Xrb$
(in other words, translate a string from S-language to T-language).
However, there might be multiple such T-strings, in which case we speak of \emph{ambiguity}.

    \begin{Xdef}
    T-strings $x$ and $y$ are \emph{ambiguous} iff $x \!\neq\! y$ and $S(x) \Xeq S(y)$.
    \end{Xdef}

We can define equivalence relation $\simeq$ on the T-language: let $x \simeq y \Leftrightarrow S(x) \Xeq S(y)$.
Under this relation each equivalence class with more than one element forms a maximal subset of pairwise ambiguous T-strings.

    \begin{Xdef}
    For a TRE $e$ \emph{disambiguation policy} is a strict partial order $\prec$ on $L \Xeq \XT \Xlb e \Xrb$, such that
    for each subset of pairwise ambiguous T-strings
    it is total ($\forall$ ambiguous $x, y \Xin L$: either $x \prec y$ or $y \prec x$),
    and the minimal T-string in this subset exists ($\exists x \Xin L: \forall y \Xin L \mid$ ambiguous $x, y: x \prec y$).
    \end{Xdef}

We will return to disambiguation in section \ref{section_disambiguation}.
\\

In practice obtaining submatch results in a form of a T-string is inconvenient.
A more practical representation is the \emph{tag value function} used by Laurikari: 
a separate list of offsets in the input string for each tag.
Tag value functions can be trivially reconstructed from T-strings.
However, the two representations are not equivalent;
in particular, tag value functions have a weaker notion of ambiguity and fail to capture ambiguity in some TRE, as shown below.
Therefore we use T-strings as a primary representation
and convert them to tag value functions after disambiguation.

    \begin{Xdef}\label{tagvalfun}
    \emph{Decomposition} of a T-string $x \Xeq \gamma_1 \dots \gamma_n$
    is a \emph{tag value function} $H: T \rightarrow (\YN \cup \{ 0, \varnothing \})^*$
    that maps each tag to a string of offsets in $S(x)$:
    $H(t) \Xeq \varphi^t_1 \dots \varphi^t_n$, where:
    $$\varphi^t_i = \begin{cases}
            \varnothing &\text{if } \gamma_i \Xeq \Xbar{t} \\[-0.5em]
            |S(\gamma_1 \dots \gamma_i)| &\text{if } \gamma_i \Xeq t \\[-0.5em]
            \epsilon &\text{otherwise}
        \end{cases}$$
    \end{Xdef}

Negative submatches have no exact offset: they can be attributed to any point on the alternative path of TRE.
We use a special value $\varnothing$ to represent them
(it is semantically equivalent to negative infinity).
\\

For example, for a T-string $x \Xeq \alpha 1 2 \beta 2 \beta \alpha 1 \Xbar{2}$,
possibly denoted by TRE $(\alpha 1 (2 \beta)^*)^*$, we have
$S(x) \Xeq \alpha \beta \beta \alpha$ and tag value function:
    $$H(t) \Xeq \begin{cases}
        1 \, 4 &\text{if } t \Xeq 1 \\[-0.5em]
        1 \, 2 \, \varnothing &\text{if } t \Xeq 2
    \end{cases}$$

Decomposition is irreversible in general:
even if we used a real offset instead of $\varnothing$, we no longer know the relative order of tags with equal offsets.
For example, TRE $(1 (3 \, 4)^{1,3} 2)^{2}$,
which may represent POSIX RE \texttt{(()\{1,3\})\{2\}},
denotes ambiguous T-strings $x \Xeq 1 3 4 3 4 2 1 3 4 2$ and $y \Xeq 1 3 4 2 1 3 4 3 4 2$.
According to POSIX, first iteration has higher priority than the second one,
and repeated empty match, if optional, should be avoided, therefore $y \prec x$.
However, both $x$ and $y$ decompose to the same tag value function:
    $$H(t) \Xeq \begin{cases}
        0 \, 0 &\text{if } t \Xeq 1 \\[-0.5em]
        0 \, 0 &\text{if } t \Xeq 2 \\[-0.5em]
        0 \, 0 \, 0 &\text{if } t \Xeq 3\\[-0.5em]
        0 \, 0 \, 0 &\text{if } t \Xeq 4
    \end{cases}$$

Moreover, consider another T-string $z \Xeq 1 3 4 2 1 3 4 3 4 3 4 2$ denoted by this RE.
By the same reasoning $z \prec x$ and $y \prec z$.
However, comparison of tag value functions cannot yield the same result
(since $x$, $y$ have the same tag value function and $z$ has a different one).
In practice this doesn't cause disambiguation errors
as long as the minimal T-string corresponds to the minimal tag value function,
but in general the order is different.
\\

Decomposition can be computed incrementally in a single left-to-right pass over the T-string:
$\alpha_i$ in definition \ref{untag} and
$\varphi_i^t$ in definition \ref{tagvalfun} depend only on $\gamma_j$ such that $j \!\leq\! i$.

\section{From TRE to automata}\label{section_tnfa}

Both S-language and T-language of the given TRE are regular,
and in this perspective submatch extraction reduces to the problem of translation between regular languages.
The class of automata capable of performing such translation is known as \emph{finite state transducers (FST)} (see e.g. [Ber13], page 68).
TNFA, as defined by Laurikari in \cite{Lau01}, is a nondeterministic FST
that decomposes output strings into tag value functions
and then applies disambiguation.
Our definition is different in the following aspects.
First, we apply disambiguation \emph{before} decomposition
(for the reasons discussed in the previous section).
Second, we do not consider disambiguation policy as an attribute of TNFA:
the same TNFA can be simulated with different policies, though not always efficiently.
Third, we include information about TRE structure in the form of \emph{prioritized} $\epsilon$-transitions:
it is used by some disambiguation policies.
Finally, we add negative tagged transitions.

    \begin{Xdef}
    \emph{Tagged Nondeterministic Finite Automaton (TNFA)}
    is a structure $(\Sigma, T, P, Q, F, q_0, \Delta)$, where:
    \begin{itemize}
    \setlength{\parskip}{0.5em}
        \item[] $\Sigma$ is a finite set of symbols (\emph{alphabet})
        \item[] $T$ is a finite set of \emph{tags}
        \item[] $P$ is a finite set of \emph{priorities}
        \item[] $Q$ is a finite set of \emph{states}
        \item[] $F \subseteq Q$ is the set of \emph{final} states
        \item[] $q_0 \in Q$ is the \emph{initial} state

        \item[] $\Delta \Xeq \Delta^\Sigma \sqcup \Delta^\epsilon$ is the \emph{transition} relation, where:
        \begin{itemize}
            \item[] $\Delta^\Sigma \subseteq Q \times \Sigma \times \{\epsilon\} \times Q$
            \item[] $\Delta^\epsilon \subseteq Q \times (P \cup \{\epsilon\}) \times (T \cup \Xbar{T} \cup \{\epsilon\}) \times Q$
        \end{itemize}


        and all $\epsilon$-transitions from the same state have different priorities:
        $\forall (x, r, \epsilon, y), (\widetilde{x}, \widetilde{r}, \epsilon, \widetilde{y}) \Xin \Delta^\epsilon:
        x \Xeq \widetilde{x} \wedge y \Xeq \widetilde{y} \Rightarrow r \!\neq\! \widetilde{r}$.
    \end{itemize}
    \end{Xdef}

TNFA construction is similar to Thompson NFA construction,
except for priorities and generalized repetition.
For the given TRE $e$ over $\Sigma$, $T$, the corresponding TNFA is $\XN(e) \Xeq (\Sigma, T, \{0, 1\}, Q, \{ y \}, x, \Delta)$,
where $(Q, x, y, \Delta) \Xeq \XF(\XX(e))$ and $\XF$ is defined as follows:
    \begin{align*}
        \XF(\emptyset) &= (\{ x, y \}, x, y, \emptyset) \\
        \XF(\epsilon) &= (\{ x, y \}, x, y, \{ (x, \epsilon, \epsilon, y) \}) \\
        \XF(\alpha) &= (\{ x, y \}, x, y, \{ (x, \alpha, \epsilon, y) \}) \\
        \XF(t) &= (\{ x, y \}, x, y, \{ (x, \epsilon, t, y) \}) \\
        \XF(e_1 | e_2) &= \XF(e_1) \cup \XF(e_2) \\
        \XF(e_1 e_2) &= \XF(e_1) \cdot \XF(e_2) \\
        \XF(e^{n,\infty}) &= \XF(e)^{n,\infty} \\
        \XF(e^{n,m}) &= \XF(e)^{n, m}
    \end{align*}
    \begin{align*}
        F_1 \cup F_2 &= (Q, x, y, \Delta) \\
        \text{where }
            & (Q_1, x_1, y_1, \Delta_1) = F_1 \\
            & (Q_2, x_2, y_2, \Delta_2) = F_2 \\
            & Q = Q_1 \cup Q_2 \cup \{ x, y \} \\
            & \Delta = \Delta_1 \cup \Delta_2 \cup \{ \\
                & \qquad (x, 0, \epsilon, x_1), (y_1, \epsilon, \epsilon, y), \\
                & \qquad (x, 1, \epsilon, x_2), (y_2, \epsilon, \epsilon, y) \}
    \end{align*}
\begin{Xfig}
\includegraphics[width=0.3\linewidth]{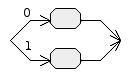}\\*
\captionof{figure}{Automata union.}
\end{Xfig}
    \begin{align*}
        F_1 \cdot F_2 &= (Q, x_1, y_2, \Delta) \\
        \text{where }
            & (Q_1, x_1, y_1, \Delta_1) = F_1 \\
            & (Q_2, x_2, y_2, \Delta_2) = F_2 \\
            & Q = Q_1 \cup Q_2  \\
            & \Delta = \Delta_1 \cup \Delta_2 \cup \{ (y_1, \epsilon, \epsilon, x_2) \}
    \end{align*}
\begin{Xfig}
\includegraphics[width=0.25\linewidth]{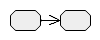}\\*
\captionof{figure}{Automata product.}
\end{Xfig}
    \begin{align*}
        F^{n,\infty} &= (Q, x_1, y_{n+1}, \Delta) \\
        \text{where }
            & \{(Q_i, x_i, y_i, \Delta_i)\}_{i=1}^n = \{F, \dots, F\} \\
            & Q = \bigcup\nolimits_{i=1}^n Q_i \cup \{ y_{n+1} \} \\
            & \Delta = \bigcup\nolimits_{i=1}^n \Delta_i
                \cup \{(y_i, \epsilon, \epsilon, x_{i+1})\}_{i=1}^{n\!-\!1} \\
                & \hphantom{\hspace{2em}}
                    \cup \{ (y_n, 0, \epsilon, x_n), (y_n, 1, \epsilon, y_{n+1}) \}
    \end{align*}
\begin{Xfig}
\includegraphics[width=0.55\linewidth]{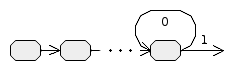}\\*
\captionof{figure}{Unbounded repetition of automata.}
\end{Xfig}
    \begin{align*}
        F^{n,m} &= (Q, x_1, y_m, \Delta) \\
        \text{where }
            & \{(Q_i, x_i, y_i, \Delta_i)\}_{i=1}^m = \{F, \dots, F\} \\
            & Q = \bigcup\nolimits_{i=1}^m Q_i \\
            & \Delta = \bigcup\nolimits_{i=1}^m \Delta_i
                \cup \{(y_i, \epsilon, \epsilon, x_{i+1})\}_{i=1}^{n\!-\!1} \\
                & \hphantom{\hspace{2em}}
                    \cup \{(y_i, 0, \epsilon, x_{i+1}), (y_i, 1, \epsilon, y_m)\}_{i=n}^{m\!-\!1}
    \end{align*}
\begin{Xfig}
\includegraphics[width=0.9\linewidth]{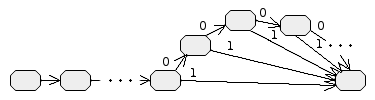}\\*
\captionof{figure}{Bounded repetition of automata.}
\end{Xfig}

The above construction of TNFA has certain properties that will be used in subsequent sections.

\begin{Xobs}\label{obs_tnfa_states}
We can partition all TNFA states into three disjoint subsets:
\begin{enumerate}
    \item states that have outgoing transitions on symbols;
    \item states that have outgoing $\epsilon$-transitions;
    \item states without outgoing transitions (including the final state);
\end{enumerate}
This statement can be proved by induction on the structure of TNFA:
automata for atomic TRE $\emptyset$, $\epsilon$, $\alpha$, $t$ obviously satisfy it;
compound automata $F_1 \cup F_2$, $F_1 \cdot F_2$, $F^{n,\infty}$ and $F^{n,m}$
do not violate it:
they only add outgoing $\epsilon$-transitions to those states that have no outgoing transitions,
and their final state is either a new state without outgoing transitions,
or final state of one of the subautomata.
\end{Xobs}

\begin{Xobs}\label{obs_tnfa_repeat}
For repetition automata $F^{n,\infty}$ and $F^{n,m}$
the number of iterations uniquely determines the order of subautomata traversal:
by construction subautomaton corresponding to $(i \!+\! 1)$-th iteration is only reachable
from the one corresponding to $i$-th iteration
(in case of unbounded repetition it may be the same subautomaton).
\end{Xobs}

    \begin{Xdef}
    A \emph{path} in TNFA $(\Sigma, T, P, Q, F, q_0, \Delta)$ is a sequence of transitions
    $\{(q_i, \alpha_i, a_i, \widetilde{q}_i)\}_{i=1}^n \subseteq \Delta$, where $n \!\geq\! 0$
    and $\widetilde{q}_i \Xeq q_{i+1} \; \forall i \Xeq \overline{1,n-1}$.
    \end{Xdef}

    \begin{Xdef}
    Path $\{(q_i, \alpha_i, a_i, \widetilde{q}_i)\}_{i=1}^n$ in TNFA $(\Sigma, T, P, Q, F, q_0, \Delta)$ is \emph{accepting}
    if either $n \Xeq 0 \wedge q_0 \Xin F$ or $n\!>\!0 \wedge q_1 \Xeq q_0 \wedge \widetilde{q}_n \Xin F$.
    \end{Xdef}

    \begin{Xdef}
    Every path $\pi \Xeq \{(q_i, \alpha_i, a_i, \widetilde{q}_i)\}_{i=1}^n$
    in TNFA $(\Sigma, T, P, Q, F, q_0, \Delta)$
    \emph{induces} an S-string, a T-string and a string over $P$ called \emph{bitcode}:
    \begin{align*}
    \XS(\pi) &= \alpha_1 \dots \alpha_n \\
    \XT(\pi) &= \alpha_1 \gamma_1 \dots \alpha_n \gamma_n
    &&\gamma_i = \begin{cases}
            a_i &\text{if } a_i \Xin T \cup \Xbar{T} \\[-0.5em]
            \epsilon &\text{otherwise}
        \end{cases} \\[-0.5em]
    \XB(\pi) &= \beta_1 \dots \beta_n
    &&\beta_i = \begin{cases}
            a_i &\text{if } a_i \Xin P \\[-0.5em]
            \epsilon &\text{otherwise}
        \end{cases}
    \end{align*}
    \end{Xdef}

    \begin{Xdef}
    TNFA $\XN$ \emph{transduces} S-string $s$ to a T-string $x$, denoted $s \xrightarrow{\XN} x$
    if $s \Xeq S(x)$ and there is an accepting path $\pi$ in $\XN$, such that $\XT(\pi) \Xeq x$.
    \end{Xdef}

The set of all S-strings that are transduced to some T-string is the \emph{input language} of TNFA;
likewise, the set of all transduced T-strings is the \emph{output language} of TNFA.
It is easy to see that for every TRE $e$ the input language of TNFA $\XN(e)$ equals to its S-language $\XS \Xlb e \Xrb$
and the output language of $\XN(e)$ equals to its T-language $\XT \Xlb e \Xrb$
(proof is by induction on the structure of TRE and by construction of TNFA).
\\

The simplest way to simulate TNFA is as follows.
Starting from the initial state, trace all possible paths that match the input string; record T-strings along each path.
When the input string ends, paths that end in a final state are accepting;
choose the one with the minimal T-string (with respect to disambiguation policy).
Convert the resulting T-string into tag value function.
At each step the algorithm maintains a set of \emph{configurations} $(q, x)$ that represent current paths:
$q$ is TNFA state and $x$ is the induced T-string.
The efficiency of this algorithm depends on the implementation of $closure$, which is discussed in the next section.
\\

    \begin{algorithm}[H] \DontPrintSemicolon \SetKwProg{Fn}{}{}{} \SetAlgoInsideSkip{medskip}
    \Fn {$\underline{transduce((\Sigma, T, P, Q, F, q_0, T, \Delta), \alpha_1 \dots \alpha_n)} \smallskip$} {
        $X \Xset closure(\{ (q_0, \epsilon) \}, F, \Delta)$ \;
        \For {$i \Xeq \overline{1,n}$} {
            $Y \Xset reach(X, \Delta, \alpha_i)$ \;
            $X \Xset closure(Y, F, \Delta)$ \;
        }
        $x \Xset min_\prec\{ x \mid (q, x) \Xin X \wedge q \Xin F \}$ \;
        \Return $H(x)$ \;
    }
    \end{algorithm}

    \begin{algorithm}[H] \DontPrintSemicolon \SetKwProg{Fn}{}{}{} \SetAlgoInsideSkip{medskip}
    \Fn {$\underline{reach(X, \Delta, \alpha)} \smallskip$} {
        \Return $\{ (p, x \alpha) \mid (q, x) \Xin X \wedge (q, \alpha, \epsilon, p) \Xin \Delta \}$
    }
    \end{algorithm}

\section{Tagged $\epsilon$-closure}\label{section_closure}

The most straightforward implementation of $closure$ (shown below)
is to simply gather all possible non-looping $\epsilon$-paths.
Note that we only need paths that end in the final state
or paths which end state has outgoing transitions on symbols:
all other paths will be dropped by $reach$ on the next simulation step.
Such states are called \emph{core states}; they belong to subsets 1 or 3 in observation \ref{obs_tnfa_states}.
\\

    \begin{algorithm}[H] \DontPrintSemicolon \SetKwProg{Fn}{}{}{} \SetAlgoInsideSkip{medskip}
    \Fn {$\underline{closure(X, F, \Delta)} \smallskip$} {
        empty $stack$, $result \Xset \emptyset$ \;
        \For {$(q, x) \Xin X:$} {
            $push(stack, (q, x))$ \;
        }
        \While {$stack$ is not empty} {
            $(q, x) \Xset pop(stack)$ \;
            $result \Xset result \cup \{(q, x)\}$ \;
            \ForEach {outgoing arc $(q, \epsilon, \chi, p) \Xin \Delta$} {
                \If {$\not \exists (\widetilde{p}, \widetilde{x})$ on stack $: \widetilde{p} \Xeq p$} {
                    $push(stack, (p, x \chi))$ \;
                }
            }
        }
        \Return $\{ (q, x) \Xin result \mid core(q, F, \Delta) \}$ \;
    }
    \end{algorithm}

    \begin{algorithm}[H] \DontPrintSemicolon \SetKwProg{Fn}{}{}{} \SetAlgoInsideSkip{medskip}
    \Fn {$\underline{core(q, F, \Delta)} \smallskip$} {
        \Return $q \Xin F \vee \exists \alpha, p: (q, \alpha, \epsilon, p) \Xin \Delta$
    }
    \end{algorithm}

Since there might be multiple paths between two given states,
the number of different paths may grow up exponentially in the number of TNFA states.
If we prune paths immediately as they arrive at the same TNFA state,
we could keep the number of active paths at any point of simulation bounded by the number of TNFA states.
However, this puts a restriction on disambiguation policy:
it must allow to compare ambiguous T-strings by their ambiguous prefixes.
We call such policy \emph{prefix-based};
later we will show that both POSIX and leftmost greedy policies have this property.

    \begin{Xdef}
    Paths
    $\pi_1 \Xeq \{(q_i, \alpha_i, a_i, \widetilde{q}_i)\}_{i=1}^n$ and
    $\pi_2 \Xeq \{(p_i, \beta_i, b_i, \widetilde{p}_i)\}_{i=1}^m$
    are \emph{ambiguous} if their start and end states coincide: $q_1 \Xeq p_1$, $\widetilde{q}_n \Xeq \widetilde{p}_m$
    and their induced T-strings $\XT(\pi_1)$ and $\XT(\pi_2)$ are ambiguous.
    \end{Xdef}

    \begin{Xdef}
    Disambiguation policy for TRE $e$ is \emph{prefix-based}
    if it can be extended on the set of ambiguous prefixes of T-strings in $\XT \Xlb e \Xrb$,
    so that for any ambiguous paths $\pi_1 $, $\pi_2 $ in TNFA $\XN \Xlb e \Xrb$
    and any common suffix $\pi_3$ the following holds:
    $\XT(\pi_1) \prec \XT(\pi_2) \Leftrightarrow \XT(\pi_1 \pi_3) \prec \XT(\pi_2 \pi_3)$.
    \end{Xdef}

The problem of closure construction can be expressed in terms of single-source shortest-path problem
in directed graph with cycles and mixed (positive and negative) arc weights.
(We assume that all initial closure states are connected to one imaginary ``source'' state).
Most algorithms for solving shortest-path problem have the same basic structure (see e.g. \cite{Cor09}, chapter 24):
starting with the source node, repeatedly scan nodes;
for each scanned node apply \emph{relaxation} to all outgoing arcs;
if path to the given node has been improved, schedule it for further scanning.
Such algorithms are based on the \emph{optimal substructure} principle:
any prefix of the shortest path is also a shortest path.
In our case tags do not map directly to weights and T-strings are more complex than distances, but direct mapping is not necessary:
optimal substructure principle still applies if the disambiguation policy is prefix-based,
and relaxation can be implemented via T-string comparison and extension of T-string along the given transition.
Also, we assume absence of epsilon-loops with ``negative weight'',
which is quite reasonable for any disambiguation policy.
Laurikari gives the following algorithm for closure construction (see Algorithm 3.4 in \cite{Lau01}):
\\

    \begin{algorithm}[H] \DontPrintSemicolon \SetKwProg{Fn}{}{}{} \SetAlgoInsideSkip{medskip}
    \Fn {$\underline{closure \Xund laurikari(X, F, \Delta)} \smallskip$} {
        empty $deque$, $result(q) \equiv \bot$ \;
        $indeg \Xset indegree(X, \Delta)$ \;
        $count \Xset indeg$ \;
        \For {$(q, x) \Xin X$} {
            $relax(q, x, result, deque, count, indeg)$ \;
        }
        \While {$deque$ is not empty} {
            $q \Xset pop \Xund front (deque)$ \;
            \ForEach {outgoing arc $(q, \epsilon, \chi, p) \Xin \Delta$} {
                $x \Xset result(q) \chi$ \;
                $relax(p, x, result, deque, count, indeg)$ \;
            }
        }
        \Return $\{ (q, x) \mid x \Xeq result(q) \wedge core(q, F, \Delta) \}$
    }
    \end{algorithm}

    \begin{algorithm}[H] \DontPrintSemicolon \SetKwProg{Fn}{}{}{} \SetAlgoInsideSkip{medskip}
    \Fn {$\underline{relax(q, x, result, deque, count, indeg)} \smallskip$} {
        \If {$x \prec result(q)$} {
            $result(q) \Xset x$ \;
            $count(p) \Xset count(p) - 1$ \;
            \If {$count(p) \Xeq 0$} {
                $count(p) \Xset indeg(p)$ \;
                $push \Xund front (deque, q)$ \;
            } \Else {
                $push \Xund back (deque, q)$ \;
            }
        }
    }
    \end{algorithm}

    \begin{algorithm}[H] \DontPrintSemicolon \SetKwProg{Fn}{}{}{} \SetAlgoInsideSkip{medskip}
    \Fn {$\underline{indegree(X, \Delta)} \smallskip$} {
        empty $stack$, $indeg(q) \equiv 0$ \;
        \For {$(q, x) \Xin X$} {
            $push(stack, q)$ \;
        }
        \While {$stack$ is not empty} {
            $q \Xset pop(stack)$ \;
            \If {$indeg(q) \Xeq 0$} {
                \ForEach {outgoing arc $(q, \epsilon, \chi, p) \Xin \Delta$} {
                    $push(stack, p)$ \;
                }
            }
            $indeg(q) \Xset indeg(q) + 1$ \;
        }
        \Return $indeg$ \;
    }
    \end{algorithm}

We will refer to the above algorithm as LAU.
The key idea of LAU is to reorder scanned nodes so that ancestors are processed before their descendants.
This idea works well for acyclic graphs: scanning nodes in topological order yields a linear-time algorithm \cite{Cor09} (chapter 24.2),
so we should expect that LAU also has linear complexity on acyclic graphs.
However, the way LAU decrements in-degree is somewhat odd: decrement only happens if relaxation was successful,
while it seems more logical to decrement in-degree every time the node is encountered.
Another deficiency is that nodes with zero in-degree may occur in the middle of the queue,
while the first node does not necessarily have zero in-degree.
These observations lead us to a modification of LAU, which we call LAU1
(all the difference is in $relax$ procedure):
\\

    \begin{algorithm}[H] \DontPrintSemicolon \SetKwProg{Fn}{}{}{} \SetAlgoInsideSkip{medskip}
    \Fn {$\underline{relax(q, x, result, deque, count, indeg)} \smallskip$} {
        \If {$count(q) \Xeq 0$} {
            $count(q) \Xset indeg(q)$ \;
        }
        $count(p) \Xset count(p) - 1$ \;

        \If {$count(p) \Xeq 0$ and $p$ is on $deque$} {
            $remove (deque, p)$ \;
            $push \Xund front (deque, p)$ \;
        }

        \If {$x \prec result(q)$} {
            $result(q) \Xset x$ \;
            \If {$q$ is not on $deque$} {
                \If {$count(q) \Xeq 0$} {
                    $push \Xund front (deque, q)$ \;
                } \Else {
                    $push \Xund back (deque, q)$ \;
                }
            }
        }
    }
    \end{algorithm}

Still for graphs with cycles worst-case complexity of LAU and LAU1 is unclear;
usually algorithms that schedule nodes in LIFO order (e.g. Pape-Levit) have exponential complexity \cite{SW81}.
However, there is another algorithm also based on the idea of topological ordering,
which has $O(nm)$ worst-case complexity and $O(n + m)$ complexity on acyclic graphs
(where $n$ is the number of nodes and $m$ is the number of edges).
It is the GOR1 algorithm described in \cite{GR93}
(the version listed here is one of the possible variations of the algorithm):
\\

    \begin{algorithm}[H] \DontPrintSemicolon \SetKwProg{Fn}{}{}{} \SetAlgoInsideSkip{medskip}
    \Fn {$\underline{closure \Xund goldberg \Xund radzik(X, F, \Delta)} \smallskip$} {
        empty stacks $topsort$, $newpass$ \;
        $result(q) \equiv \bot$ \;
        $status(q) \equiv \mathit{OFFSTACK}$ \;
        \For {$(q, x) \Xin X$} {
            $relax(q, x, result, topsort)$ \;
        }
        \While {$topsort$ is not empty} {
            \While {$topsort$ is not empty} {
                $q \Xset pop(topsort)$ \;

                \If {$status(q) \Xeq \mathit{TOPSORT}$} {
                    $push(newpass, n)$ \;
                } \ElseIf {$status(q) \Xeq \mathit{NEWPASS}$} {
                    $status(q) \Xset \mathit{TOPSORT}$ \;
                    $push(topsort, q)$ \;
                    $scan(q, result, topsort)$ \;
                }
            }
            \While {$newpass$ is not empty} {
                $q \Xset pop(newpass)$ \;
                $scan(q, result, topsort)$ \;
                $status(q) \Xset \mathit{OFFSTACK}$ \;
            }
        }
        \Return $\{ (q, x) \mid x \Xeq result(q) \wedge core(q, F, \Delta) \}$ \;
    }
    \end{algorithm}

    \begin{algorithm}[H] \DontPrintSemicolon \SetKwProg{Fn}{}{}{} \SetAlgoInsideSkip{medskip}
    \Fn {$\underline{scan(q, result, topsort)} \smallskip$} {
        \ForEach {outgoing arc $(q, \epsilon, \chi, p) \Xin \Delta$} {
            $x \Xset result(q) \chi$ \;
            $relax(p, x, result, topsort)$ \;
        }
    }
    \end{algorithm}

    \begin{algorithm}[H] \DontPrintSemicolon \SetKwProg{Fn}{}{}{} \SetAlgoInsideSkip{medskip}
    \Fn {$\underline{relax(q, x, result, topsort)} \smallskip$} {
        \If {$x \prec result(q)$} {
            $result(q) \Xset x$ \;
            \If {$status(q) \neq \mathit{TOPSORT}$} {
                $push(topsort, q)$ \;
                $status(q) \Xset \mathit{NEWPASS}$ \;
            }
        }
    }
    \end{algorithm}

In order to better understand all three algorithms and compare their behavior on various classes of graphs
I used the benchmark suite described in \cite{CGR96}.
I implemented LAU, LAU1 and the above version of GOR1;
source codes are freely available in \cite{Tro17} and open for suggestions and bug fixes.
The most important results are as follows.
On Acyc-Neg family (acyclic graphs with mixed weights)
LAU is non-linear and significantly slower,
while LAU1 and GOR1 are both linear and LAU1 scans each node exactly once.
On Grid-NHard and Grid-PHard families (graphs with cycles designed to be hard for algorithms that exploit graph structure)
both LAU and LAU1 are very slow (though approximation suggests polynomial, not exponential fit),
while GOR1 is fast.
On other graph families all three algorithms behave quite well;
it is strange that LAU is fast on Acyc-Pos family, while being so slow on Acyc-Neg family.
See also \cite{NPX99}: they study two modifications of GOR1, one of which is very close to LAU1,
and conjecture (without a proof) that worst-case complexity is exponential.

\end{multicols}

\begin{Xfig}
\includegraphics[width=\linewidth]{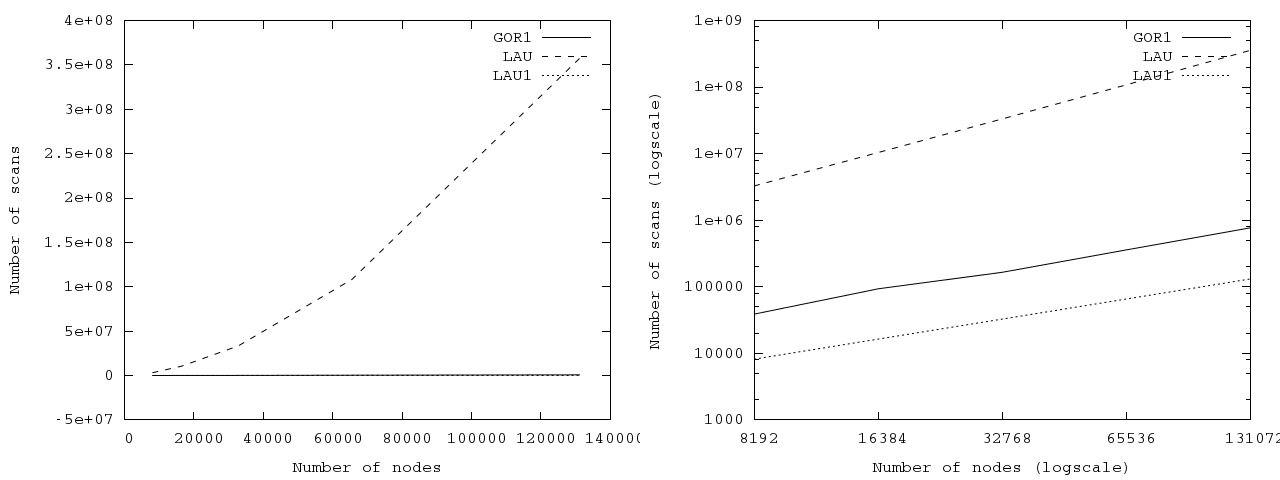}\\*
\captionof{figure}{Behavior of LAU, LAU1 and GOR1 on Acyc-Neg family of graphs.\\
Left: normal scale, right: logarithmic scale on both axes.}
\end{Xfig}

\begin{Xfig}
\includegraphics[width=\linewidth]{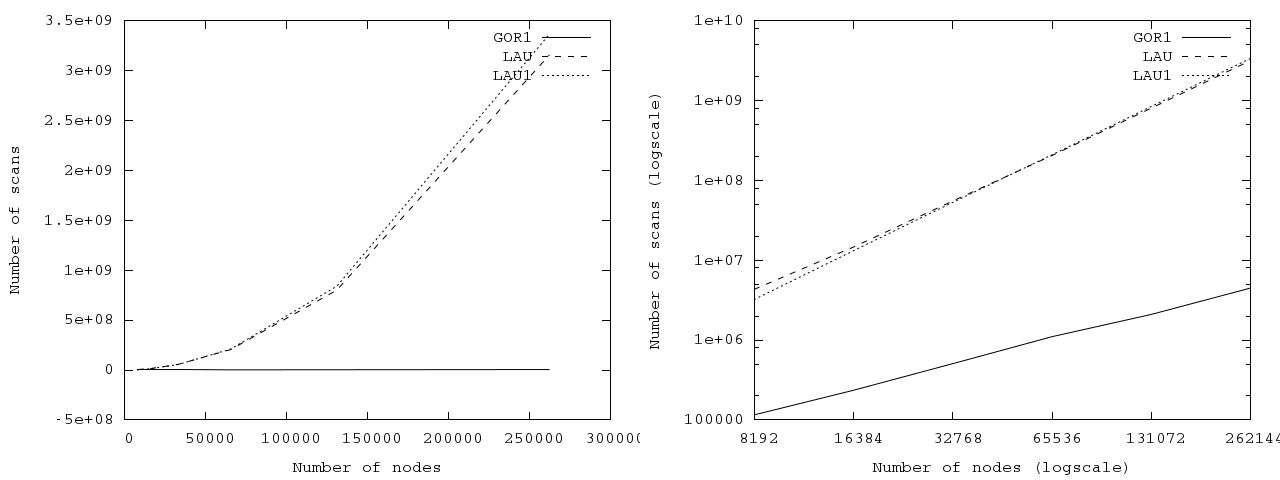}\\*
\captionof{figure}{Behavior of LAU, LAU1 and GOR1 on Grid-Nhard family of graphs.\\
Left: normal scale, right: logarithmic scale on both axes.}
\end{Xfig}

\begin{multicols}{2}


\section{Disambiguation}\label{section_disambiguation}

In section \ref{section_tagged_extension} we defined disambiguation policy as strict partial order on the T-language of the given TRE.
In practice T-language may be very large or infinite
and explicit listing of all ambiguous pairs is not an option; we need a comparison algorithm.
There are two main approaches: structure-based and value-based.
Structure-based disambiguation is guided by the order of operators in RE; tags play no role in it.
Value-based disambiguation is the opposite: it is defined in terms of maximization/minimization of certain tag parameters.
As a consequence, it has to deal with conflicts between different tags ---
a complication that never arises for structure-based approach.
Moreover, in value-based disambiguation different tags may have different rules and relations.
Below is a summary of two real-world policies supported by RE2C:
\\

\begin{itemize}
    \setlength{\parskip}{0.5em}
    \item Leftmost greedy.
        Match the longest possible prefix of the input string
        and take the \emph{leftmost path} through RE that corresponds to this prefix:
        in unions prefer left alternative, in iterations prefer re-iterating.

    \item POSIX.
        Each subexpression including the RE itself should match as early as possible
        and span as long as possible, while not violating the whole match.
        Subexpressions that start earlier in RE have priority over those starting later.
        Empty match is considered longer than no match;
        repeated empty match is allowed only for non-optional repetitions.
    \\
\end{itemize}

As we have already seen, a sufficient condition for efficient TNFA simulation is that the policy is prefix-based.
What about determinization?
In order to construct TDFA we must be able to fold loops:
if there is a nonempty loop in TNFA, determinization must eventually construct a loop in TDFA
(otherwise it won't terminate).
To do this, determinization must establish \emph{equivalence} of two TDFA states.
From disambiguation point of view equivalence means that all ambiguities stemming from one state
are resolved in the same way as ambiguities stemming from the other.
However, we cannot demand exact coincidence of all state attributes engaged in disambiguation:
if there is loop, attributes in one state are extensions of those in the other state (and hence not equal).
Therefore we need to abstract away from absolute paths and define ``ambiguity shape'' of each state: relative order on all its configurations.
Disambiguation algorithm must be defined in terms of relative paths, not absolute paths.
Then we could compare states by their orders.
If disambiguation policy can be defined in this way, we call it \emph{foldable}.
\\

In subsequent sections we will formally define both policies in terms of comparison of ambiguous T-strings
and show that each policy is prefix-based and foldable.

\subsection*{Leftmost greedy}

Leftmost greedy policy was extensively studied by many authors; we will refer to \cite{Gra15}, as their setting is very close to ours.
We can define it as lexicographic order on the set of all bitcodes corresponding to ambiguous paths
(see \cite{Gra15}, definition 3.25).
Let $\pi_1$, $\pi_2$ be two ambiguous paths which induce T-strings $x \Xeq \XT(\pi_1)$, $y \Xeq \XT(\pi_2)$
and bitcodes $a \Xeq \XB(\pi_1)$, $b \Xeq \XB(\pi_2)$.
Then $x \prec y$ iff $\prec_{lexicographic} (a, b)$:
\\

    \begin{algorithm}[H] \DontPrintSemicolon \SetKwProg{Fn}{}{}{} \SetAlgoInsideSkip{medskip}
    \Fn {$\underline{\prec_{lexicographic} (a_1 \dots a_n, b_1 \dots b_m)} \smallskip$} {
        \For {$i \Xeq \overline{1, min(n, m)}$} {
            \lIf {$a_i \!\neq\! b_i$} { \Return $a_i \!<\! b_i$  }
        }
        \Return $n \!<\! m$ \;
    }
    \end{algorithm}

This definition has one caveat: the existence of minimal element is not guaranteed for TRE that contain $\epsilon$-loops.
For example, TNFA for $\epsilon^+$ has infinitely many ambiguous paths with bitcodes
of the form $\widehat{0}^n \widehat{1}$, where $n \!\geq\! 0$,
and each bitcode is lexicographically less than the previous one.
Paths that contain $\epsilon$-loops are called \emph{problematic} (see \cite{Gra15}, definition 3.28).
If we limit ourselves to non-problematic paths (e.g. by cancelling loops in $\epsilon$-closure),
then the minimal element exists and bitcodes are well-ordered.

\begin{XLem}\label{lemma_bitcodes}
Let $\Pi$ be a set of TNFA paths that start in the same state, induce the same S-string and end in a core state
(e.g. the set of active paths on each step of TNFA simulation).
Then the set of bitcodes induced by paths in $\Pi$ is prefix-free
(compare with \cite{Gra15}, lemma 3.1).
\\[0.5em]
\textbf{Proof.}
Consider paths $\pi_1$ and $\pi_2$ in $\Pi$
and suppose that $\XB(\pi_1)$ is a prefix of $\XB(\pi_2)$.
Then $\pi_1$ must be a prefix of $\pi_2$: otherwise there is a state where $\pi_1$ and $\pi_2$ diverge,
and by TNFA construction all outgoing transitions from this state have different priorities,
which contradicts the equality of bitcodes.
Let $\pi_2 \Xeq \pi_1 \pi_3$.
Since $\XS(\pi_1) \Xeq \XS(\pi_2)$, and since $\XS(\rho\sigma) \Xeq \XS(\rho)\XS(\sigma)$ for arbitrary path $\rho\sigma$,
it must be that $\XS(\pi_3) \Xeq \epsilon$.
The end state of $\pi_2$ is a core state: by observation \ref{obs_tnfa_states} it has no outgoing $\epsilon$-transitions.
But the same state is also the start state of $\epsilon$-path $\pi_3$, therefore $\pi_3$ is an empty path and $\pi_1 \Xeq \pi_2$.
$\square$
\end{XLem}

From lemma \ref{lemma_bitcodes} it easily follows that leftmost greedy disambiguation is prefix-based.
Consider ambiguous paths $\pi_1$, $\pi_2$ and arbitrary suffix $\pi_3$,
and let $\XB(\pi_1) \Xeq a$, $\XB(\pi_2) \Xeq b$, $\XB(\pi_3) \Xeq c$.
Note that $\XB(\rho\sigma) \Xeq \XB(\rho)\XB(\sigma)$ for arbitrary path $\rho\sigma$,
therefore $\XB(\pi_1\pi_3) \Xeq ac$ and $\XB(\pi_2\pi_3) \Xeq bc$.
If $a \Xeq b$, then $ac \Xeq bc$.
Otherwise, without loss of generality let $a \prec_{lexicographic} b$: since $a$, $b$ are prefix-free, $ac \prec_{lexicographic} bc$
(compare with \cite{Gra15}, lemma 2.2).
\\

From lemma \ref{lemma_bitcodes} it also follows that leftmost greedy disambiguation is foldable:
prefix-free bitcodes can be compared incrementally on each step of simulation.
We define ``ambiguity shape'' of TDFA state as lexicographic order on bitcodes of all paths represented by configurations
(compare with \cite{Gra15}, definition 7.14).
The number of different weak orderings of $n$ elements is finite, therefore determinization terminates
(this number equals $\sum_{k=0}^n \Xstirling{n}{k} k!$, also known as the \emph{ordered Bell number}).
Order on configurations is represented with ordinal numbers assigned to each configuration.
Ordinals are initialized to zero and then updated on each step of simulation by comparing bitcodes.
Bitcodes are compared incrementally:
first, by ordinals calculated on the previous step, then by bitcode fragments added by the $\epsilon$-closure.
\\

    \begin{algorithm}[H] \DontPrintSemicolon \SetKwProg{Fn}{}{}{} \SetAlgoInsideSkip{medskip}
    \Fn {$\underline{\prec_{leftmost \Xund greedy} ((n, a), (m, b))} \smallskip$} {
        \lIf {$n \!\neq\! m$} {\Return $n \!<\! m$}
        \Return $\prec_{lexicographic} (a, b)$ \;
    }
    \end{algorithm}

    \begin{algorithm}[H] \DontPrintSemicolon \SetKw{Let}{let} \SetKw{Und}{undefined} \SetKwProg{Fn}{}{}{} \SetAlgoInsideSkip{medskip}
    \Fn {$\underline{ordinals (\{(q_i, o_i, x_i)\}_{i=1}^n)} \smallskip$} {
        $\{(p_i, B_i)\} \Xset $ sort $\{(i, (o_i, x_i))\}$ by second component using $\prec_{leftmost \Xund greedy}$ \;
        \Let $o_{p_1}(t) \Xeq 0$, $ord \Xset 0$ \;
        \For {$i \Xeq \overline{2, n}$} {
            \lIf {$B_{i-1} \!\neq\! B_i$} {$ord \Xset ord \!+\! 1$}
            \Let $o_{p_i}(t) \Xeq ord$ \;
        }
        \Return $\{(q_i, o_i, x_i)\}_{i=1}^n$ \;
    }
    \end{algorithm}

In practice explicit calculation of ordinals and comparison of bitcodes is not necessary:
if we treat TDFA states as ordered sets,
sort TNFA transitions by their priority
and define $\epsilon$-closure as a simple depth-first search,
then the first path that arrives at any state would be the leftmost.
This approach is taken in e.g. \cite{Kar14}.
Since tags are not engaged in disambiguation,
we can use paired tags that represent capturing parentheses, or just standalone tags --- this makes no difference with leftmost greedy policy.

\subsection*{POSIX}

POSIX policy is defined in \cite{POSIX}; \cite{Fow03} gives a comprehensible interpretation of it.
We will give a formal interpretation in terms of tags;
it was first described by Laurikari in \cite{Lau01}, but the key idea should be absolutely attributed to Kuklewicz \cite{Kuk07}.
He never fully formalized his algorithm, and our version slightly deviates from the informal description,
so all errors should be attributed to the author of this paper.
Fuzz-testing RE2C against Regex-TDFA revealed a couple of rare bugs in submatch extraction in Regex-TDFA,
but for the most part the two implementations agree
(see section \ref{section_tests_and_benchmarks} for details).
\\

POSIX disambiguation is defined in terms of \emph{subexpressions} and \emph{subpatterns}:
subexpression is a parenthesized sub-RE and subpattern is a non-parenthesized sub-RE.
Submatch extraction applies only to subexpressions, but disambiguation applies to both:
subpatterns have ``equal rights'' with subexpressions.
For simplicity we will now assume that all sub-RE are parenthesized;
later in this section we will discuss the distinction in more detail.
\\

POSIX disambiguation is hierarchical:
each subexpression has a certain \emph{priority} relative to other subexpressions,
and disambiguation algorithm must consider subexpressions in the order of their priorities.
Therefore we will start by enumerating all subexpressions of the given RE according to POSIX standard:
outer ones before inner ones and left ones before right ones.
Enumeration is done by rewriting RE $e$ into an \emph{indexed RE} (IRE): a pair $(i, \widetilde{e})$,
where $i$ is the index and $\widetilde{e}$ mirrors the structure of $e$,
except that each sub-IRE is an indexed pair rather than a RE.
For example, RE $a^* (b | \epsilon)$ corresponds to IRE
$(1,
    (2, (3, a)^*)
    (4, (5, b) | (6, \epsilon))
)$.
Enumeration operator $\XI$ is defined below:
it transforms a pair $(e, i)$ into a pair $(\widetilde{e}, j)$,
where $e$ is a RE, $i$ is the start index, $\widetilde{e}$ is the resulting IRE and $j$ is the next free index.
    \begin{align*}
        \XI(\emptyset, i) &= ((i, \emptyset), i \!+\! 1) \\
        \XI(\epsilon, i) &= ((i, \epsilon), i \!+\! 1) \\
        \XI(\alpha, i) &= ((i, \alpha), i \!+\! 1) \\
        \XI(e_1 | e_2, i) &= ((i, \widetilde{e_1} | \widetilde{e_2}), k) \\
            \text{where }
            & (\widetilde{e_1}, j) \Xeq \XI(e_1, i \!+\! 1),
            \; (\widetilde{e_2}, k) \Xeq \XI(e_2, j \!+\! 1) \\
        \XI(e_1 e_2, i) &= ((i, \widetilde{e_1} \widetilde{e_2}), k) \\
            \text{where }
            & (\widetilde{e_1}, j) \Xeq \XI(e_1, i \!+\! 1),
            \; (\widetilde{e_2}, k) \Xeq \XI(e_2, j \!+\! 1) \\
        \XI(e^{n,m}, i) &= ((i, \widetilde{e}^{n,m}), j) \\
            \text{where } & (\widetilde{e}, j) \Xeq \XI(e, i \!+\! 1)
    \end{align*}

Now that the order on subexpressions is defined, we can
rewrite IRE into TRE by rewriting each indexed subexpression $(i, e)$
into tagged subexpression $t_1 e \, t_2$, where $t_1 \Xeq 2i \!-\! 1$ is the \emph{start tag}
and $t_2 \Xeq 2i$ is the \emph{end tag}.
If $e$ is a repetition subexpression, then $t_1$ and $t_2$ are called \emph{orbit} tags.
TRE corresponding to the above example is
$(1\,
    3\, (5\, a \,6)^* \,4\,
    7\, (9\, b \,10 | 11\, \epsilon \,12) 8\,
2)$.
\\

According to POSIX, each subexpression should start as early as possible and span as long as possible.
In terms of tags this means that the position of start tag is \emph{minimized},
while the position of the end tag is \emph{maximized}.
Subexpression may match several times, therefore one tag may occur multiple times in the T-string.
Obviously, orbit tags may repeat;
non-orbit tags also may repeat provided that they are nested in a repetition subexpression.
For example, TRE $(1 \, (3 \, (5 \, a \, 6 | 7 \, b \, 8) \, 4)^* \, 2)$ that corresponds to POSIX RE \texttt{(a|b)*}
denotes T-string $1\, 3\, 5\, a\, 6\, \Xbar{7}\, \Xbar{8}\, 4\, 3\, 5\, a\, 6\, \Xbar{7}\, \Xbar{8}\, 4\, 2$
(corresponding to S-string $aa$),
in which orbit tags $3$ and $4$ occur twice, as well as non-orbit tags $5$, $6$, $7$ and $8$.
Each occurrence of tag has a corresponding \emph{offset}:
either $\varnothing$ (for negative tags), or the number of preceding symbols in the S-string.
The sequence of all offsets is called \emph{history}:
for example, tag $3$ has history $0 \, 1$ and tag $7$ has history $\varnothing \, \varnothing$.
Each history consists of one or more \emph{subhistories}:
longest subsequences of offsets not interrupted by tags of subexpressions with higher priority.
In our example tag $3$ has one subhistory $0 \, 1$, while tag $7$ has two subhistories $\varnothing$ and $\varnothing$.
Non-orbit subhistories contain exactly one offset (possibly $\varnothing$);
orbit subhistories are either $\varnothing$, or may contain multiple non-$\varnothing$ offsets.
Histories can be reconstructed from T-strings as follows:

    \begin{algorithm}[H] \DontPrintSemicolon \SetKwProg{Fn}{}{}{} \SetAlgoInsideSkip{medskip}
    \Fn {$\underline{history(a_1 \dots a_n, t)} \smallskip$} {
        $i \Xset 1, \; j \Xset 1, \; pos \Xset 0$ \;
        \While {$true$} {
            \While {$i \leq n$ and $a_i \!\not\in\! \{t, \Xbar{t}\}$} {
                \lIf {$a_i \Xin \Sigma$} {$pos \Xset pos \!+\! 1$}
                $i \Xset i \!+\! 1$ \;
            }
            \While {$i \leq n$ and $a_i \!\not\in\! hightags(t)$} {
                \lIf {$a_i \Xin \Sigma$}   {$pos \Xset pos \!+\! 1$}
                \lIf {$a_i \Xeq t$}        {$A_j \Xset A_j pos$}
                \lIf {$a_i \Xeq \Xbar{t}$} {$A_j \Xset A_j \varnothing$}
                $i \Xset i \!+\! 1$ \;
            }
            \lIf {$i \!>\! n$} {break}
            $j \Xset j \!+\! 1$ \;
        }
        \Return $A_1 \dots A_j$ \;
    }
    \end{algorithm}

    \begin{algorithm}[H] \DontPrintSemicolon \SetKwProg{Fn}{}{}{} \SetAlgoInsideSkip{medskip}
    \Fn {$\underline{hightags(t)} \smallskip$} {
        \Return $\{ u, \Xbar{u} \mid u < 2 \lceil t / 2 \rceil \!-\! 1 \}$ \;
    }
    \end{algorithm}

Due to the hierarchical nature of POSIX disambiguation, if comparison reaches $i$-th subexpression,
it means that all enclosing subexpressions have already been compared and their tags coincide.
Consequently the number of subhistories of tags $2i - 1$ and $2i$ in the compared T-strings must be equal.
\\

If disambiguation is defined on T-string prefixes, then the last subhistory may be incomplete.
In particular, last subhistory of start tag may contain one more offset than last subhistory of end tag.
In this case we assume that the missing offset is $\infty$, as it must be greater than any offset in the already matched S-string prefix.
\\

Disambiguation algorithm for TRE with $N$ subexpressions is defined as comparison of T-strings $x$ and $y$:

    \begin{algorithm}[H] \DontPrintSemicolon \SetKw{Let}{let} \SetKw{Und}{undefined} \SetKwProg{Fn}{}{}{} \SetAlgoInsideSkip{medskip}
    \Fn {$\underline{\prec_{POSIX}(x, y)} \smallskip$} {
        \For {$t \Xeq \overline{1, N}$} {
            $A_1 \dots A_n \Xset history(x, 2t \!-\! 1)$ \;
            $C_1 \dots C_n \Xset history(x, 2t)$ \;
            $B_1 \dots B_n \Xset history(y, 2t \!-\! 1)$ \;
            $D_1 \dots D_n \Xset history(y, 2t)$ \;
            \For {$i \Xeq \overline{1, n}$} {
                \Let $a_1 \dots a_m \Xeq A_i$, $b_1 \dots b_k \Xeq B_i$ \;
                \Let $c_1 \dots c_{\widetilde{m}} \Xeq C_i$, $d_1 \dots d_{\widetilde{k}} \Xeq D_i$ \;
                \lIf {$\widetilde{m} \!<\! m$} {$c_m \Xset \infty$}
                \lIf {$\widetilde{k} \!<\! k$} {$d_k \Xset \infty$}
                \For {$j \Xeq \overline{1, min(m, k)}$} {
                    \lIf {$a_j \!\neq\! b_j$} {\Return $a_j \!<\! b_j$}
                    \lIf {$c_j \!\neq\! d_j$} {\Return $c_j \!>\! d_j$}
                }
                \lIf {$m \!\neq\! k$} {\Return $m \!<\! k$}
            }
        }
        \Return $false$ \;
    }
    \end{algorithm}

It's not hard to show that $\prec_{POSIX}$ is prefix-based.
Consider $t$-th iteration of the algorithm and let $s \Xeq 2t \!-\! 1$ be the start tag,
$history(x, s) \Xeq A_1 \dots A_n$ and $history(y, s) \Xeq B_1 \dots B_n$.
The value of each offset depends only on the number of preceding $\Sigma$-symbols,
therefore for an arbitrary suffix $z$ we have:
$history(xz, s) \Xeq A_1 \dots A_{n-1} A'_n C_1 \dots C_n$ and
$history(yz, s) \Xeq B_1 \dots B_{n-1} B'_n C_1 \dots C_n$,
where $A'_n \Xeq A_n c_1 \dots c_m$, $B'_n \Xeq B_n c_1 \dots c_m$.
The only case when $z$ may affect comparison is when $m \!\geq\! 1$ and one history is a proper prefix of the other:
$A_i \Xeq B_i$ for all $i \Xeq \overline{1,n-1}$ and (without loss of generality) $B_n \Xeq A_n b_1 \dots b_k$.
Otherwise either histories are equal, or comparison terminates before reaching $c_1$.
Let $d_1 \dots d_{k+m} \Xeq b_1 \dots b_k c_1 \dots c_m$.
None of $d_j$ can be $\varnothing$, because $n$-th subhistory contains multiple offsets.
Therefore $d_j$ are non-decreasing and $d_j \!\leq\! c_j$ for all $j \Xeq \overline{1, m}$.
Then either $d_j \!<\! c_j$ at some index $j \!\leq\! m$, or $A'_n$ is shorter than $B'_n$; in both cases comparison is unchanged.
The same reasoning holds for the end tag.
\\

It is less evident that $\prec_{POSIX}$ is foldable:
the rest of this chapter is a long and tiresome justification of Kuklewicz algorithm
(with a couple of modifications and ideas by the author).
\\

First, we simplify $\prec_{POSIX}$.
It makes a lot of redundant checks:
for adjacent tags the position of the second tag is fixed on the position of the first tag.
In particular, comparison of the start tags $a_j$ and $b_j$ is almost always redundant.
Namely, if $j \!>\! 1$, then $a_j$ and $b_j$ are fixed on $c_{j-1}$ and $d_{j-1}$, which have been compared on the previous iteration.
If $j \Xeq 1$, then $a_j$ and $b_j$ are fixed on some higher-priority tag which has already been checked, unless $t \Xeq 1$.
The only case when this comparison makes any difference is when $j \Xeq 1$ and $t \Xeq 1$:
the very first position of the whole match.
In order to simplify further discussion we will assume that the match is anchored;
otherwise one can handle it as a special case of comparison algorithm.
The simplified algorithm looks like this:
\\

    \begin{algorithm}[H] \DontPrintSemicolon \SetKwProg{Fn}{}{}{} \SetAlgoInsideSkip{medskip}
    \Fn {$\underline{\prec_{POSIX}(x, y)} \smallskip$} {
        \For {$t \Xeq \overline{1, N}$} {
            $A_1 \dots A_n \Xset history(x, 2t)$ \;
            $B_1 \dots B_n \Xset history(y, 2t)$ \;
            \For {$i \Xeq \overline{1, n}$} {
                \lIf {$A_i \!\neq\! B_i$} {\Return $A_i \prec_{subhistory} B_i$}
            }
        }
        \Return $false$ \;
    }
    \end{algorithm}

    \begin{algorithm}[H] \DontPrintSemicolon \SetKwProg{Fn}{}{}{} \SetAlgoInsideSkip{medskip}
    \Fn {$\underline{\prec_{subhistory} (a_1 \dots a_n, b_1 \dots b_m)} \smallskip$} {
        \For {$i \Xeq \overline{1, min(n, m)}$} {
            \lIf {$a_i \!\neq\! b_i$} {\Return $a_i \!>\! b_i$}
        }
        \Return $n \!<\! m$ \;
    }
    \end{algorithm}

Next, we explore the structure of ambiguous paths that contain multiple subhistories
and show that (under certain conditions) such paths can be split into ambiguous subpaths,
one per each subhistory.

\begin{XLem}\label{lemma_path_decomposition}
Let $e$ be a POSIX TRE and suppose that the following conditions are satisfied:
\begin{enumerate}
    \item $a$, $b$ are ambiguous paths in TNFA $\XN(e)$ that induce T-strings $x \Xeq \XT(a)$, $y \Xeq \XT(b)$
    \item $t$ is a tag such that $history(x, t) \Xeq A_1 \dots A_n$, $history(y, t) \Xeq B_1 \dots B_n$
    \item for all $u \!<\! t$: $history(x, u) \Xeq history(y, u)$
        (tags with higher priority agree)
\end{enumerate}
Then $a$ and $b$ can be decomposed into path segments $a_1 \dots a_n$, $b_1 \dots b_n$,
such that for all $i \!\leq\! n$ subpaths $a_i$, $b_i$ have common start and end states and
contain subhistories $A_i$, $B_i$ respectively:
$history(\XT(a_1 \dots a_i), t)$ $\Xeq$ $A_1 \dots A_i$,
$history(\XT(b_1 \dots b_i), t)$ $\Xeq$ $B_1 \dots B_i$.
\\[0.5em]
\textbf{Proof.}
Proof is by induction on $t$ and relies on the construction of TNFA given in section \ref{section_tnfa}.
Induction basis is $t \Xeq 1$ and $t \Xeq 2$ (start and end tags of the topmost subexpression): let $n \Xeq 1$, $a_1 \Xeq a$, $b_1 \Xeq b$.
Induction step: suppose that lemma is true for all $u \!<\! t$,
and for $t$ the conditions of lemma are satisfied.
Let $r$ be the start tag of a subexpression in which $t$ is immediately enclosed.
Since $r \!<\! t$, the lemma is true for $r$ by inductive hypothesis;
let $c_1 \dots c_m$, $d_1 \dots d_m$ be the corresponding path decompositions.
Each subhistory of $t$ is covered by some subhistory of $r$ (by definition $history$ doesn't break at lower-priority tags),
therefore decompositions $a_1 \dots a_n$, $b_1 \dots b_n$ can be constructed as a refinement of $c_1 \dots c_m$, $d_1 \dots d_m$.
If $r$ is a non-orbit tag, each subhistory of $r$ covers exactly one subhistory of $t$
and the refinement is trivial: $n \Xeq m$, $a_i \Xeq c_i$, $b_i \Xeq d_i$.
Otherwise, $r$ is an orbit tag and single subhistory of $r$ may contain multiple subhistories of $t$.
Consider path segments $c_i$ and $d_i$:
since they have common start and end states, and since they cannot contain tagged transitions with higher-priority tags,
both must be contained in the same subautomaton of the form $F^{k,l}$.
This subautomaton itself consists of one or more subautomata for $F$ each starting with an $r$-tagged transition;
let the start state of each subautomaton be a breaking point in the refinement of $c_i$ and $d_i$.
By observation \ref{obs_tnfa_repeat} the number of iterations through $F^{k,l}$ uniquely determines the order of subautomata traversal.
Since $history(x, r) \Xeq history(y, r)$, the number of iterations is equal and
therefore breaking points coincide.
$\square$
\end{XLem}

Lemma \ref{lemma_path_decomposition} has the following implication.
Suppose that during simulation we prune ambiguous paths immediately as they transition to to the same state,
and suppose that
at $p$-th step of simulation we are comparing histories $A_1 \dots A_n$, $B_1 \dots B_n$ of some tag.
Let $j \!\leq\! n$ be the greatest index such that all offsets in $A_1 \dots A_j$, $B_1 \dots B_j$ are less than $p$
(it must be the same index for both histories because higher-priority tags coincide).
Then $A_i \Xeq B_i$ for all $i \!\leq\! j$:
by lemma \ref{lemma_path_decomposition} $A_1 \dots A_j$, $B_1 \dots B_j$
correspond to subpaths which start and end states coincide;
these subpaths are either equal, or ambiguous, in which case they must have been compared on some previous step of the algorithm.
This means that we only need to compare $A_{j+1} \dots A_n$ and $B_{j+1} \dots B_n$.
Of them only $A_{j+1}$ and $B_{j+1}$ may have offsets less than $p$:
all other subhistories belong to current $\epsilon$-closure;
the last pair of subhistories $A_n$, $B_n$ may be incomplete.
Therefore we only need to remember $A_{j+1}$, $B_{j+1}$ from the previous simulation step,
and we only need to pass $A_n$, $B_n$ to the next step.
In other words, between simulation steps we need only the last subhistory for each tag.
\\

Now we can define ``ambiguity shape'' of TDFA state:
we define it as a set of orders, one per tag, on the last subhistories of this tag in this state.
As with leftmost greedy policy, the number of different orders is finite
and therefore determinization terminates.
In fact, comparison only makes sense for subhistories that correspond to ambiguous paths (or path prefixes),
and only in case when higher-priority tags agree.
We do not know in advance which prefixes will cause ambiguity on subsequent steps,
therefore some comparisons may be meaningless:
we impose total order on a set which is only partially ordered.
However, meaningless comparisons do not affect valid comparisons, and they do not cause disambiguation errors: their results are never used.
At worst they can prevent state merging.
Kuklewicz suggests to group orbit subhistories by their \emph{base offset}
(position of start tag on the first iteration) prior to comparison.
However, experiments with such grouping revealed no effect on state merging,
and for simplicity we abandon the idea of partial ordering.

\begin{Xdef}
Subhistories of the given tag are \emph{comparable} if they correspond to prefixes of ambiguous paths
and all higher-priority tags agree.
\end{Xdef}

\begin{XLem}\label{lemma_orbit_subhistories}
Comparable orbit subhistories can be compared incrementally with $\prec_{subhistory}$.
\\[0.5em]
\textbf{Proof.}
Consider subhistories $A$, $B$ at some step of simulation and let $A \prec_{subhistory} B$.
We will show that comparison result will not change on subsequent steps, when new offsets are added to $A$ and $B$.
First, note that $\varnothing$ can be added only on the first step of comparison:
negative orbit tags correspond to the case of zero iterations,
and by TNFA construction for $F^{0,m}$ they are reachable by $\epsilon$-transitions from the initial state,
but not from any other state of this subautomaton.
Second, note that non-$\varnothing$ offsets increase with each step.
Based on these two facts and the definition of $\prec_{subhistory}$, the proof is trivial by induction on the number of steps.
$\square$
\end{XLem}

\begin{XLem}\label{lemma_nonorbit_end_subhistories}
Comparable non-orbit subhistories can be compared incrementally with $\prec_{subhistory}$
in case of end tags, but not in case of start tags.
\\[0.5em]
\textbf{Proof.}
Non-orbit subhistories consist of a single offset (either $\varnothing$ or not),
and ambiguous paths may discover it at different steps.
Incremental comparison with $\prec_{subhistory}$ is correct in all cases except one:
when $\varnothing$ is discovered at a later step than non-$\varnothing$.
\\[0.5em]
For start tags it is sufficient to show an example of such case.
Consider TRE $1 (3 \, a \, 4 | 5 \, a \, 6) 2$ that corresponds to POSIX RE \texttt{(a)|(a)}
and denotes ambiguous T-strings $x \Xeq 1 \, 3 \, a \, 4 \, \Xbar{5} \, \Xbar{6} \, 2$
and $y \Xeq 1 \, 5 \, a \, 6 \, \Xbar{3} \, \Xbar{4} \, 2$.
Subhistory of start tag $3$ in $y$ changes from $\epsilon$ on the first step (before consuming \texttt{a})
to $\varnothing$ on the second step (after consuming \texttt{a}),
while subhistory in $x$ remains $0$ on both steps.
\\[0.5em]
For end tags we will show that the faulty case is not possible:
comparable subhistories must add $\varnothing$ at the same step as non-$\varnothing$.
Consider non-orbit end tag $t$.
Non-$\varnothing$ and $\varnothing$ must stem from different alternatives of a union subexpression $e_1 | e_2$,
where $e_1$ contains $t$ and $e_2$ does not.
Since subhistories of $t$ are comparable, $e_1$ cannot contain higher-priority tags:
such tags would be negated in $e_2$ and comparison would stop before $t$.
Consequently, $e_1$ itself must be the subexpression that ends with $t$.
By construction of TNFA for $e_1 | e_2$
all paths through it contain a single $t$-tagged transition at the very end (either positive or negative).
Therefore both $\varnothing$ and non-$\varnothing$ must be discovered at the same step when ambiguous paths join.
$\square$
\end{XLem}

This asymmetry between start and end tags in caused by inserting negative tags
at the \emph{end} of alternative branches;
if we inserted them at the \emph{beginning},
then non-orbit tags would also have the property that $\varnothing$ belongs to the first step of comparison.
Inserting negative tags at the end has other advantage: it effectively delays the associated operations,
which should result in more efficient programs.
Since our disambiguation algorithm ignores start tags,
we can use the same comparison algorithm for all subhistories.
Alternatively one can compare non-orbit tags using simple maximization/minimization strategy:
if both last offsets of the given tag belong to the $\epsilon$-closure, they are equal;
if only one of them belongs to the $\epsilon$-closure, it must be greater than the other one;
otherwise the result of comparison on the previous step should be used.
\\

Orders are represented with vectors of ordinal numbers (one per tag) assigned to each configuration.
Ordinals are initialized to zero and updated on each step of simulation by comparing last subhistories.
Subhistories are compared using ordinals from the previous step and T-string fragments added by the $\epsilon$-closure.
Ordinals are assigned in decreasing order, so that they can be compared like offsets:
greater values have higher priority.
\\

    \begin{algorithm}[H] \DontPrintSemicolon \SetKw{Let}{let} \SetKw{Und}{undefined} \SetKwProg{Fn}{}{}{} \SetAlgoInsideSkip{medskip}
    \Fn {$\underline{ordinals (\{(q_i, o_i, x_i)\}_{i=1}^n)} \smallskip$} {
        \For {$t \Xeq \overline{1, N}$} {
            \For {$i \Xeq \overline{1, n}$} {
                $A_1 \dots A_m \Xset \epsilon \Xund history(x_i, t)$ \;
                $B_i \Xset A_m$ \;
                \lIf {$m \Xeq 1$} {$B_i \Xset o_i(t) B_i$}
            }
            \BlankLine
            $\{(p_i, C_i)\} \Xset $ sort $\{(i, B_i)\}$ by second component using inverted $\prec_{subhistory}$ \;
            \Let $o_{p_1}(t) \Xeq 0$, $ord \Xset 0$ \;
            \For {$i \Xeq \overline{2, n}$} {
                \lIf {$C_{i-1} \!\neq\! C_i$} {$ord \Xset ord \!+\! 1$}
                \Let $o_{p_i}(t) \Xeq ord$ \;
            }
        }
        \Return $\{(q_i, o_i, x_i)\}_{i=1}^n$ \;
    }
    \end{algorithm}

The $history$ algorithm is modified to handle T-string fragments added by the $\epsilon$-closure:
non-$\varnothing$ offsets are set to $\infty$, as all tags in the $\epsilon$-closure have the same offset
which is greater than any ordinal calculated on the previous step.
\\

    \begin{algorithm}[H] \DontPrintSemicolon \SetKwProg{Fn}{}{}{} \SetAlgoInsideSkip{medskip}
    \Fn {$\underline{\epsilon \Xund history (a_1 \dots a_n, t)} \smallskip$} {
        $i \Xset 1, \; j \Xset 1$ \;
        \While {$true$} {
            \While {$i \leq n$ and $a_i \!\not\in\! hightags(t)$} {
                \lIf {$a_i \Xeq t$}        {$A_j \Xset A_j \infty$}
                \lIf {$a_i \Xeq \Xbar{t}$} {$A_j \Xset A_j \varnothing$}
                $i \Xset i \!+\! 1$ \;
            }
            \lIf {$i \!>\! n$} {break}
            $j \Xset j \!+\! 1$ \;
            \While {$i \leq n$ and $a_i \!\not\in\! \{t, \Xbar{t}\}$} {
                $i \Xset i \!+\! 1$ \;
            }
        }
        \Return $A_1 \dots A_j$ \;
    }
    \end{algorithm}

Disambiguation is defined as comparison of pairs $(ox, x)$ and $(oy, y)$,
where $ox$, $oy$ are ordinals and $x$, $y$ are the added T-string fragments:
\\

    \begin{algorithm}[H] \DontPrintSemicolon \SetKwProg{Fn}{}{}{} \SetAlgoInsideSkip{medskip}
    \Fn {$\underline{\prec_{POSIX}((ox, x), (oy, y))} \smallskip$} {
        \For {$t \Xeq \overline{1, N}$} {
            $A_1 \dots A_n \Xset \epsilon \Xund history(x, 2t), \; a \Xset ox(2t)$ \;
            $B_1 \dots B_n \Xset \epsilon \Xund history(y, 2t), \; b \Xset oy(2t)$ \;
            $A_1 \Xset a A_1$ \;
            $B_1 \Xset b B_1$ \;


            \For {$i \Xeq \overline{1, n}$} {
                \lIf {$A_i \!\neq\! B_i$} {\Return $A_i \prec_{subhistory} B_i$}
            }
        }
        \Return $false$ \;
    }
    \end{algorithm}

So far we have treated all subexpressions uniformly as if they were marked for submatch extraction.
In practice most of them are not: we can reduce the amount of tags by dropping all tags in subexpressions without nested submatches
(since no other tags depend on them).
However, all the hierarchy of tags from the topmost subexpression down to each submatch must be preserved,
including \emph{fictive} tags that don't correspond to any submatch and exist purely for disambiguation purposes.
They are probably not many: POSIX RE use the same operator for grouping and submatching,
and compound expressions usually need grouping to override operator precedence,
so it is uncommon to construct a large RE without submatches.
However, fictive tags must be inserted into RE; neither Laurikari nor Kuklewicz mention it,
but both their libraries seem to do it (judging by the source code).
\\

In this respect TDFA-based matchers have an advantage over TNFA-based ones:
disambiguation happens at determinization time,
and afterwards we can erase all fictive tags -- the resulting TDFA will have no overhead.
However, if it is necessary to reduce the amount of tags at all costs (even at disambiguation time),
then fictive tags can be dropped and the algorithm modified as follows.
Each submatch should have two tags (start and end)
and repeated submatches should also have a third (orbit) tag.
Start and end tags should be maximized, if both conflicting subhistories are non-$\varnothing$;
otherwise, if only one is $\varnothing$, leftmost path should be taken;
if both are $\varnothing$, disambiguation should continue with the next tag.
Orbit tags obey the same rules as before.
The added complexity is caused by the possible absence of tags in the left part of union and concatenation.
We won't go into further details, as the modified algorithm is probably not very useful;
but an experimental implementation in RE2C passed all relevant tests in \cite{Fow03}.
Correctness proof might be based on the limitations of POSIX RE due to the coupling of groups and submatches.

\section{Determinization}\label{section_determinization}

When discussing TNFA simulation we paid little attention to tag value functions:
decomposition must wait until disambiguation, which is defined on T-strings,
and in general this means waiting until the very end of simulation.
However, since then we have studied leftmost greedy and POSIX policies more closely
and established that both are prefix-based and foldable.
This makes them suitable for determinization, but also opens possibilities for more efficient simulation.
In particular, there's no need to remember the whole T-string for each active path:
we only need ordinals and the most recent fragment added by the $\epsilon$-closure.
All the rest can be immediately decomposed into tag value function.
Consequently, we extend configurations with vectors of \emph{tag values}:
in general, each value is an offset list of arbitrary length,
but in practice values may be single offsets or anything else.
\\

Laurikari determinization algorithm has the same basic principle as the usual powerset construction (see e.g. \cite{HU90}, Theorem 2.1 on page 22):
simulation of nondeterministic automaton on all possible inputs combined with merging of equivalent states.
The most tricky part is merging: extended configuration sets are no longer equal, as they contain absolute tag values.
In section \ref{section_disambiguation} we solved similar problem with respect to disambiguation
by moving from absolute T-strings to relative ordinals.
However, this wouldn't work with tag values, as we need the exact offsets.
Laurikari resolved this predicament using \emph{references}:
he noticed that we can represent tag values as cells in ``memory'' and address each value by reference to the cell that holds it.
If states $X$ and $Y$ are equal up to renaming of references,
then we can convert $X$ to $Y$ by copying the contents of cells in $X$ to the cells in $Y$.
The number of different cells needed at each step is finite:
it is bounded by the number of tags times the number of configurations in the given state.
Therefore ``memory'' can be modeled as a finite set of \emph{registers},
which brings us to the following definition of TDFA:

    \begin{Xdef}
    \emph{Tagged Deterministic Finite Automaton (TDFA)}
    is a structure $(\Sigma, T, \YQ, \YF, Q_0, R, \delta, \zeta, \eta, \iota)$, where:
    \begin{itemize}
    \setlength{\parskip}{0.5em}
        \item[] $\Sigma$ is a finite set of \emph{symbols} (\emph{alphabet})
        \item[] $T$ is a finite set of \emph{tags}
        \item[] $\YQ$ is a finite set of \emph{states}
        \item[] $\YF \subseteq \YQ$ is the set of \emph{final} states
        \item[] $Q_0 \in \YQ$ is the \emph{initial} state
        \item[] $R$ is a finite set of \emph{registers}
        \item[] $\delta: \YQ \times \Sigma \to \YQ$ is the \emph{transition} function
        \item[] $\zeta: \YQ \times \Sigma \times R \to R \times \YB^*$ \\
            is the \emph{register update} function
        \item[] $\eta: \YF \times R \to R \times \YB^*$ \\
            is the \emph{register finalize} function
        \item[] $\iota: R \to R \times \YB^*$ \\
            is the \emph{register initialize} function
    \end{itemize}
    where $\YB$ is the boolean set $\{0,1\}$.
    \end{Xdef}

Operations on registers are associated with transitions, final states and start state,
and have the form $r_1 \Xeq r_2 b_1 \dots b_n$, where $b_1 \dots b_n$ are booleans 
$1$, $0$ denoting \emph{current position} and \emph{default value}.
For example, $r_1 \Xeq 0$ means ``set $r_1$ to default value'',
$r_1 \Xeq r_2$ means ``copy $r_2$ to $r_1$'' and
$r_1 \Xeq r_1 1 1$ means ``append current position to $r_1$ twice''.
\\

TDFA definition looks very similar to the definition of
\emph{deterministic streaming string transducer (DSST)}, described by Alur and Černý in \cite{AC11}.
Indeed, the two kinds of automata are similar and have similar applications: DSSTs are used for RE parsing in \cite{Gra15}.
However, their semantics is different: TDFA operates on tag values, while DSST operates on strings of the output language.
What is more important, DSST is \emph{copyless}:
its registers can be only \emph{moved}, not \emph{copied}.
TDFA violates this restriction, but this doesn't affect its performance as long as registers hold scalar values.
Fortunately, as we shall see, it is always possible to represent tag values as scalars.
\\

TDFA can be constructed in two slightly different ways
depending on whether we associate $\epsilon$-closure of each state with the \emph{incoming} transition,
or with all \emph{outgoing} transitions.
For the usual powerset construction it makes no difference, but things change in the presence of tagged transitions.
In the former case register operations are associated with the \emph{incoming} transition and should be executed \emph{after} it.
In the latter case they belong to each \emph{outgoing} transition and should be executed \emph{before} it,
which means that we can exploit the \emph{lookahead} symbol to filter out only the relevant part of $\epsilon$-closure:
pick only those $\epsilon$-paths which end states have transitions on the lookahead symbol.
This leaves out many useless register operations:
intuitively, we delay their application until the right lookahead symbol shows up.
However, state mapping becomes more complex:
since the operations are delayed,
their effect on each state is not reflected in configurations at the time of mapping.
In order to ensure state equivalence we must additionally demand exact coincidence of delayed operations.
\\

The two ways of constructing TDFA resemble slightly of LR(0) and LR(1) automata; we call them TDFA(0) and TDFA(1).
Indeed, we can define \emph{conflict} as a situation when tag has at least two different values in the given state.
Tags that induce no conflicts are \emph{deterministic};
the maximal number of different values per state is the tag's \emph{degree of nondeterminism}.
Accordingly, \emph{tag-deterministic} RE are those for which it is possible to build TDFA without conflicts
(also called \emph{one-pass} in \cite{Cox10}).
As with LR(0) and LR(1), many RE are tag-deterministic with respect to TDFA(1), but not TDFA(0).
Unlike LR automata, TDFA with conflicts are correct, but they can be very inefficient:
the higher tag's degree of nondeterminism, the more registers it takes to hold its values,
and the more operations are required to manage these registers.
Deterministic tags need only a single register and can be implemented without copy operations.
\\

Laurikari used TDFA(0); we study both methods and argue that TDFA(1) is better.
Determinization algorithm is defined on Figure 7;
it handles both types of automata in a uniform way.
States are sets of configurations $(q, v, o, x)$,
where $q$ is a core TNFA state, $v$ is a vector of registers that hold tag values, $o$ is the ordinal
and $x$ is the T-string of the $\epsilon$-path by which $q$ was reached.
The last component, $x$, is used only by TDFA(1), as it needs to check coincidence of delayed register operations;
for TDFA(0) it is always $\epsilon$.
During construction of $\epsilon$-closure configurations are extended to the form $(q, v, o, x, y)$,
where $y$ is the new T-string: TDFA(0) immediately applies it to tag values,
but TDFA(1) applies $x$ and delays $y$ until the next step.
Registers are allocated for all new operations:
the same register may be used on multiple outgoing transitions for operations of the same tag,
but different tags never share registers.
We assume an infinite number of vacant registers and allocate them freely, not trying to reuse old ones;
this results in a more optimization-friendly automaton.
Note also that the same set of \emph{final registers} is reused by all final states:
this simplifies tracking of final tag values.
Mapping of a newly constructed state $X$ to an existing state $Y$ checks coincidence of TNFA states, orders, delayed operations,
and constructs bijection between registers of $X$ and $Y$.
If $r_1$ in $X$ corresponds to $r_2$ in $Y$ (and they are not equal), then $r_1$ must be copied to $r_2$ on the transition to $X$
(which will become transition to $Y$ after merging).
It may happen so that $r_1$ itself is a left-hand side of an operation on this transition:
in this case we simply substitute it with $r_2$ instead of copying.
Determinization algorithm can handle both POSIX and leftmost greedy policies,
but in the latter case it can be simplified to avoid explicit calculation of ordinals, as discussed in section \ref{section_disambiguation}.

\begin{figure*}
\begin{algorithm}[H] \DontPrintSemicolon \SetKw{Let}{let} \SetKw{Und}{undefined} \SetKwProg{Fn}{}{}{} \SetAlgoInsideSkip{medskip}
\begin{multicols}{2}

    \Fn {$\underline{determinization(\XN \Xeq (\Sigma, T, Q, F, q_0, T, \Delta), \ell)} \smallskip$} {

        \tcc {initialization}
        \Let $initord(t) \Xeq 0$ \;
        \Let $initreg(t) \Xeq t$ \;
        \Let $finreg(t) \Xeq t + |T|$ \;
        \Let $maxreg \Xeq 2|T|$ \;
        \Let $newreg \equiv $ \Und \;

        \BlankLine
        \tcc {initial closure and reg-init function}
        \hangindent=1.5em\hangafter=1
        $(Q_0, regops, maxreg, newreg) \Xset closure(\XN, \ell, $
            $\{(q_0, initreg, initord, \epsilon)\}, maxreg, newreg)$ \;
        \Let $\YQ \Xeq \{ Q_0 \}$, $\YF \Xeq \emptyset$ \;
        \ForEach {$(r_1, r_2, h) \Xin regops$} {
            \Let $\iota(r_1) \Xeq (r_2, h)$ \;
        }

        \BlankLine
        \tcc {main loop}
        \While {exists unmarked state $X \Xin \YQ$} {
            mark $X$ \;

            \BlankLine
            \tcc {explore all outgoing transitions}
            \Let $newreg \equiv $ \Und \;
            \ForEach {symbol $\alpha \in \Sigma$} {
                $Y \Xset reach'(\Delta, X, \alpha)$ \;
                \hangindent=1.5em\hangafter=1
                $(Z, regops, maxreg, newreg) \Xset closure(\XN, \ell, Y, maxreg, newreg)$ \;

                \BlankLine
                \tcc {try to find mappable state}
                \If {exists $Z' \Xin \YQ$ for which $regops' \Xeq $ $map(Z', Z, T, regops) \!\neq\! $ \Und,} {
                    $(Z, regops) \Xset (Z', regops')$ \;
                } \lElse {
                    add $Z$ to $\YQ$
                }

                \BlankLine
                \tcc {transition and reg-update functions}
                \Let $\delta(X, \alpha) \Xeq Z$ \;
                \ForEach {$(r_1, r_2, h) \Xin regops$} {
                    \Let $\zeta(X, \alpha, r_1) \Xeq (r_2, h)$ \;
                }
            }

            \BlankLine
            \tcc {final state and reg-finalize function}
            \If {exists $(q, v, o, x) \Xin X \mid q \Xin F$} {
                add $X$ to $\YF$ \;
                \ForEach {tag $t \Xin T$} {
                    \Let $\eta(X, finreg(t)) \Xeq (v(t), op(x, t))$ \;
                }
            }
        }

        \BlankLine
        \Let $R \Xeq \{ 1, \dots, maxreg \}$ \;
        \Return $(\Sigma, T, \YQ, \YF, Q_0, R, \delta, \zeta, \eta, \iota)$ \;
    }
    \BlankLine
    \BlankLine

    \Fn {$\underline{op(x, t)} \smallskip$} {
        \Switch {$x$} {
            \lCase {$\epsilon$} {\Return $\epsilon$}
            \lCase {$\Xbar{t} y$} {\Return $0 \cdot op(y, t)$}
            \lCase {$t y$} {\Return $1 \cdot op(y, t)$}
            \lCase {$a y$} {\Return $op(y, t)$}
        }
    }

    \vfill\null

\columnbreak

    \Fn {$\underline{closure(\XN, lookahead, X, maxreg, newreg)} \smallskip$} {

        \tcc {construct closure and update ordinals}
        $Y \Xset \{(q, o, \epsilon) \mid (q, v, o, x) \Xin X \}$ \;
        $Y \Xset closure' (Y, F, \Delta)$ \;
        $Y \Xset ordinals (Y)$ \;
        $Z \Xset \{(q, v, \widetilde{o}, x, y) \mid (q, v, o, x) \Xin X \wedge (q, \widetilde{o}, y) \Xin Y \}$ \;

        \BlankLine
        \tcc {if TDFA(0), apply lookahead operations}
        \If {not $lookahead$} {
            $Z \Xset \{(q, v, o, y, \epsilon) \mid (q, v, o, x, y) \Xin Z \}$ \;
        }

        \BlankLine
        \tcc {find all distinct operation right-hand sides}
        \Let $newops \Xeq \emptyset$ \;
        \ForEach {configuration $(q, v, o, x, y) \Xin Z$} {
            \ForEach {tag $t \Xin T$} {
                $h \Xset op(x, t)$ \;
                \lIf {$h \!\neq\! \epsilon$} {add $(t, v(t), h)$ to $newops$}
            }
        }

        \BlankLine
        \tcc {allocate registers for new operations}
        \ForEach {$o \Xin newops$} {
            \If {$newreg(o) \Xeq $ \Und} {
                $maxreg \Xset maxreg + 1$ \;
                \Let $newreg(o) \Xeq maxreg$ \;
            }
        }

        \BlankLine
        \tcc {update registers in closure}
        \ForEach {configuration $(q, v, o, x, y) \Xin Z$} {
            \ForEach {tag $t \Xin T$} {
                $h \Xset op(x, t)$ \;
                \lIf {$h \!\neq\! \epsilon$} {\Let $v(t) \Xeq newreg(t, v(t), h)$}
            }
        }

        \BlankLine
        $X \Xset \{(q, v, o, y) \mid (q, v, o, x, y) \Xin Z \}$ \;
        $regops \Xset \{(newreg(o), r, h) | o \Xeq (t, r, h) \Xin newops\}$ \;
        \Return $(X, regops, maxreg, newreg)$ \;
    }
    \BlankLine
    \BlankLine

    \Fn {$\underline{map(X, Y, T, ops)} \smallskip$} {
        \Let $xregs(t) \Xeq \{v(t) \mid (q, v, o, x) \Xin X \}$ \;
        \Let $yregs(t) \Xeq \{v(t) \mid (q, v, o, x) \Xin Y \}$ \;

        \BlankLine

        \tcc {map one state to the other
            so that the corresponding configurations have equal TNFA states, ordinals and lookahead operations,
            and there is bijection between registers}
        \If {exists bijection $M: X \leftrightarrow Y$,
            and $\forall t \Xin T$ exists bijection $m(t): xregs(x) \leftrightarrow yregs(t)$,
            such that $\forall ((q, v, o, x), (\widetilde{q}, \widetilde{v}, \widetilde{o}, \widetilde{x})) \Xin M:$
            $q \Xeq \widetilde{q}$ and $o \Xeq \widetilde{o}$
            and $\forall t \Xin T$:
            $op(x, t) \Xeq op(\widetilde{x}, t)$
            and $(v(t), \widetilde{v}(t)) \Xin m(t)$,
            } {


            \Let $m \Xeq \bigcup_{t \in T} m(t)$ \;

            \tcc {fix target register in existing operations}
            $ops_1 \Xset \{ (a, c, h) \mid (a, b) \Xin m \wedge (b, c, h) \Xin ops \}$ \;

            \tcc {add copy operations}
            $ops_2 \Xset \{ (a, b, \epsilon) \mid (a, b) \Xin m \wedge a \!\neq\! b$
                $\hphantom{hspace{1.5em}} \wedge \nexists c, h: (b, c, h) \Xin ops \}$ \;

            \Return $ops_1 \cup ops_2$ \;
        } \lElse {
            \Return \Und
        }
    }

\end{multicols}
\end{algorithm}
\begin{center}
\caption{Determinization algorithm.\\
Functions $reach'$ and $closure'$ are exactly as
$reach$ from section \ref{section_tnfa} and $closure \Xund goldberg \Xund radzik$ from section \ref{section_closure},
except for the trivial adjustments to carry around ordinals and pass them into disambiguation procedure.
}
\end{center}
\end{figure*}

\begin{XThe}
Determinization algorithm terminates.
\\[0.5em]
\textbf{Proof.}
The proof is very similar to the one given by Laurikari in \cite{Lau00}:
we will show that for arbitrary TNFA with $t$ tags and $n$ states the number of unmappable TDFA states is finite.
Each TDFA state with $m$ configurations (where $m \!\leq\! n$) is a combination of the following components:
a set of $m$ TNFA states,
$t$ $m$-vectors of registers,
$k$ $m$-vectors of ordinals ($k \Xeq 1$ for leftmost greedy policy and $k \Xeq t$ for POSIX policy),
and an $m$-vector of T-strings.
Consider each component in turn.
First, a set of TNFA states: the number of different subsets of $n$ states is finite.
Second, a vector of registers: we assume an infinite number of registers during determinization,
but there is only a finite number of $m$-element vectors different up to bijection.
Third, a vector of ordinals: the number of different weak orderings of $m$ elements is finite.
Finally, a vector of T-strings: each T-string is induced by an $\epsilon$-path without loops,
therefore its length is bounded by the number of TNFA states,
and the number of different T-strings of length $n$ over finite alphabet of $t$ tags is finite.
$\square$
\end{XThe}

Now let's see the difference between TDFA(0) and TDFA(1) on a series of small examples.
Each example is illustrated with five pictures:
TNFA and both kinds of TDFA, each in two forms: expanded and compact.
Expanded form shows the process of determinization.
TDFA states under construction are shown as tables, where rows are configurations:
the first column is TNFA state, subsequent columns are registers used for each tag.
TDFA(1) may have additional columns for lookahead operations;
for TDFA(0) they are reflected in register versions.
Ordinals are omitted for brevity: in case of leftmost greedy policy they coincide with row indices.
Dotted states and transitions illustrate the process of mapping:
each dotted state has a transition to solid state (labeled with reordering operations).
Initializer and finalizers are also dotted;
final register versions are shown in parentheses.
Discarded ambiguous paths (if any) are shown in light gray.
Compact form shows the resulting TDFA.
Alphabet symbols on TNFA transitions are shown as ASCII codes.
TDFA transitions are labeled with numbers instead of symbols: each number represents a class of symbols
(in all the examples below number 1 corresponds to symbol \texttt{a} and number 2 to symbol \texttt{b}).
Operations are separated by forward slash ``/'' and take two forms: normal form $r_1 \Xeq r_2 b_1 \dots b_n$
and short form $r b$, which means ``set $r$ to $b$''.
Symbols $\uparrow$ and $\downarrow$ are used instead of 1 and 0 to denote \emph{current position} and \emph{default value}.
All graphs in this section are autogenerated with RE2C, so they reflect exactly the constructed automata.
By default we use leftmost greedy disambiguation, as it allows to study standalone tags and generate smaller pictures.
Note that the resulting automata are not yet optimized and use more registers than necessary.
\\

\end{multicols}

\begin{Xfig}
\includegraphics[width=0.9\linewidth]{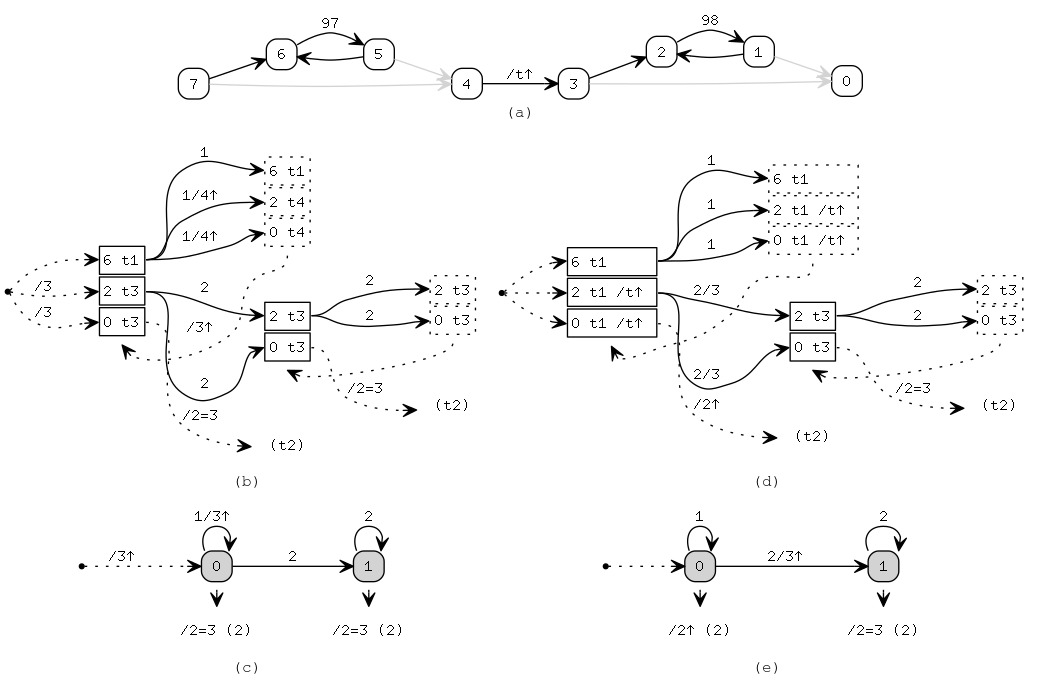}\\*
\textbf{Example 1.} $a^* 1 b^*$ (the TRE mentioned in the introduction).\\*
(a) --- TNFA, (b) --- construction of TDFA(0), (c) --- TDFA(0), (d) --- construction of TDFA(1), (e) --- TDFA(1).
\\*\medskip
This example is very simple, but it shows an important use case:
finding the edge between two non-overlapping components of the input string.
As the pictures show, TDFA(0) behaves much worse than TDFA(1):
it pulls the operation inside of loop and repeatedly rewrites tag value on each iteration,
while TDFA(1) saves it only once, when the lookahead symbol changes from \texttt{a} to \texttt{b}.
TRE is deterministic with respect to TDFA(1)
and has 2nd degree of nondeterminism with respect to TDFA(0)
(as there are at most two different registers used in each state).
\end{Xfig}

\begin{Xfig}
\includegraphics[width=0.9\linewidth]{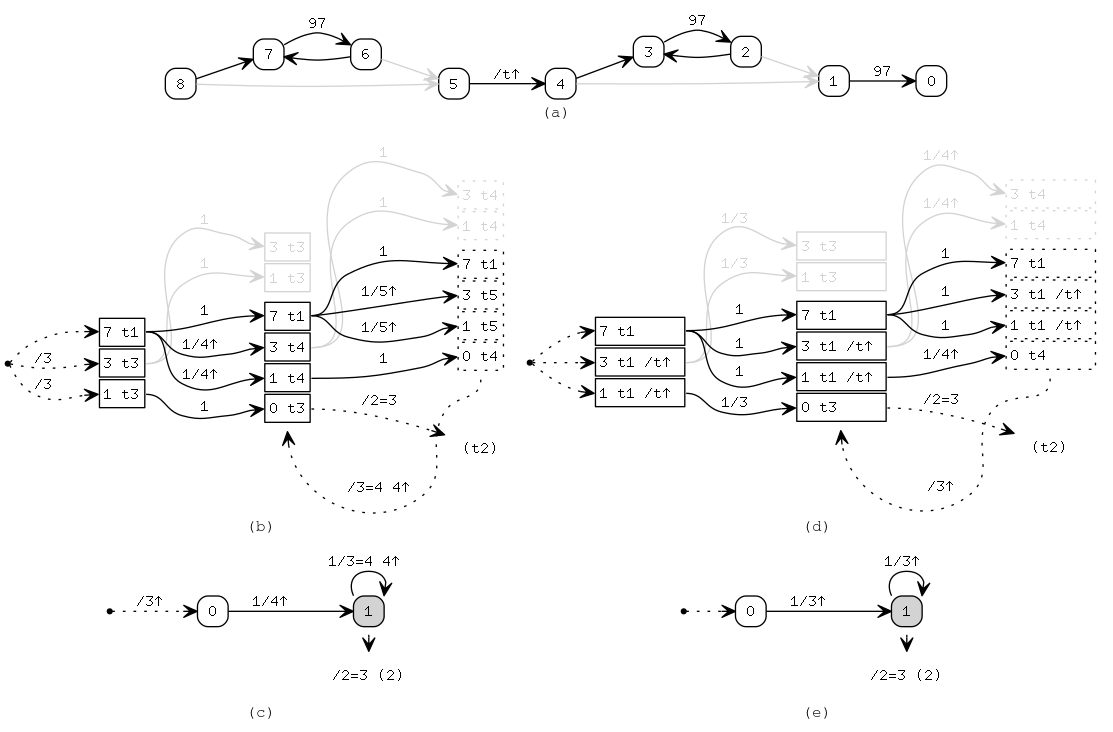}\\*
\textbf{Example 2.} $a^* 1 a^* a$ (the TRE used by Laurikari to explain his algorithm).\\*
(a) --- TNFA, (b) --- construction of TDFA(0), (c) --- TDFA(0), (d) --- construction of TDFA(1), (e) --- TDFA(1).\\*
This TRE has a modest degree of nondeterminism: 2 for TDFA(1) and 3 for TDFA(0).
Compare (c) with figure 3 from \cite{Lau00}: it is the same automaton up to a minor notational difference
(in this case leftmost greedy policy agrees with POSIX).
\end{Xfig}

\begin{Xfig}
\includegraphics[width=0.8\linewidth]{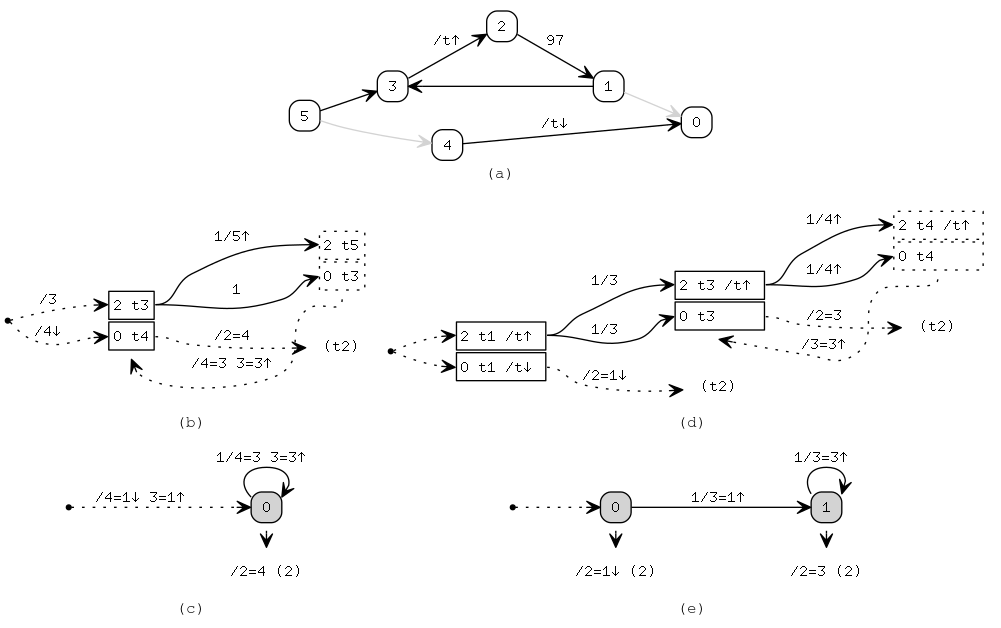}\\*
\textbf{Example 3.} $(1 a)^*$ .\\*
(a) --- TNFA, (b) --- construction of TDFA(0), (c) --- TDFA(0), (d) --- construction of TDFA(1), (e) --- TDFA(1).\\*
This example shows the typical difference between automata:
TDFA(0) has less states, but more operations; its operations are more clustered and interrelated.
Both automata record the full history of tag on all iterations.
TRE has 2nd degree nondeterminism for TDFA(0) and is deterministic for TDFA(1).
\end{Xfig}

\begin{Xfig}
\includegraphics[width=0.8\linewidth]{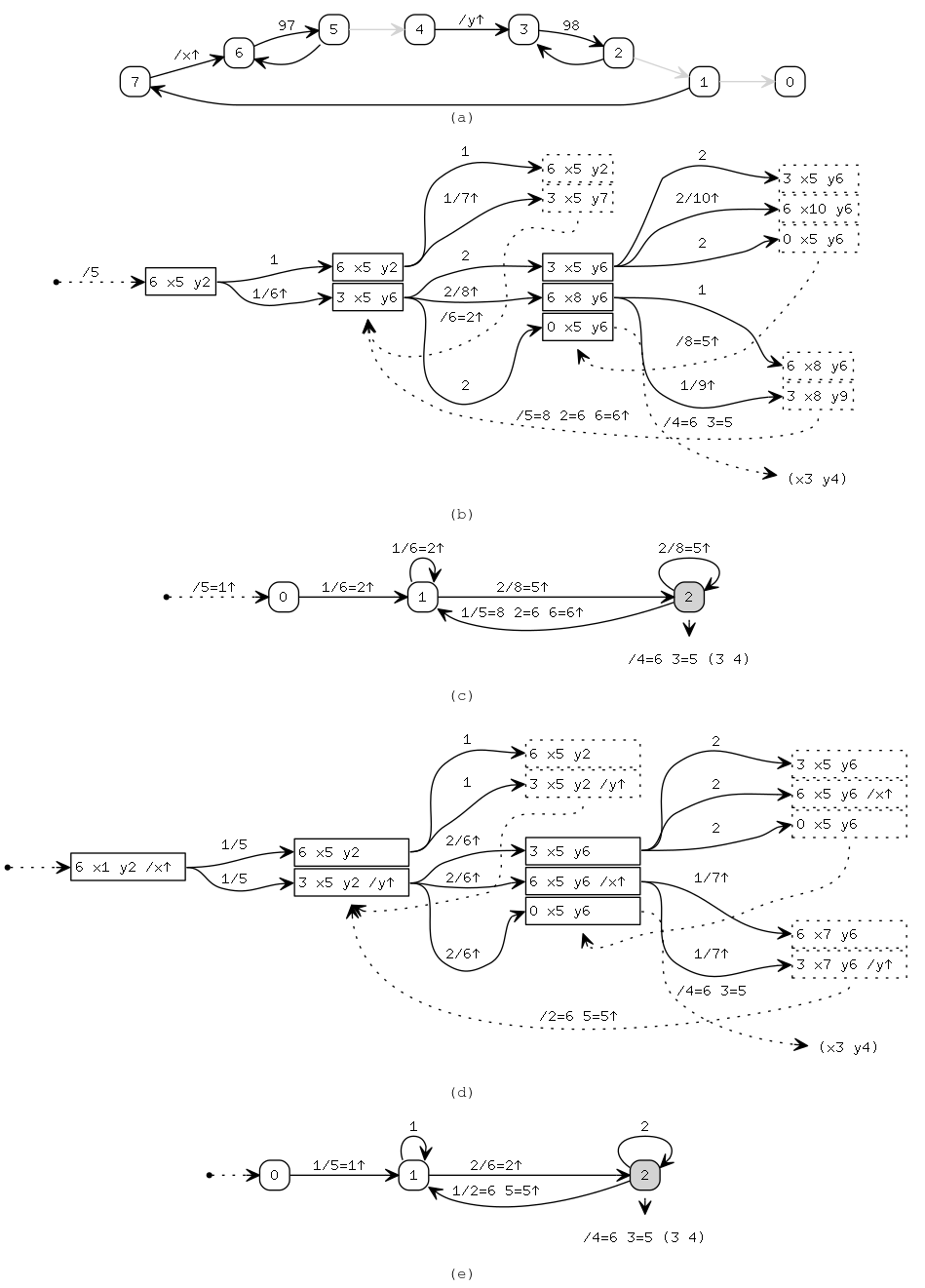}\\*
\textbf{Example 4.} $(1 a^+ 2 b^+)^+$ .\\*
(a) --- TNFA, (b) --- construction of TDFA(0), (c) --- TDFA(0), (d) --- construction of TDFA(1), (e) --- TDFA(1).\\*
Like Example 1, this example shows that TDFA(0) tends to pull operations inside of loops
and behaves much worse than hypothetical hand-written code
(only this example is bigger and gives an idea how the difference between automata changes with TRE size).
If $a^+$ and $b^+$ match multiple iterations (which is likely in practice for TRE of such form), then the difference is considerable.
Both tags have 2nd degree of nondeterminism for TDFA(0), and both are deterministic for TDFA(1).
\end{Xfig}

\begin{Xfig}
\includegraphics[width=0.9\linewidth]{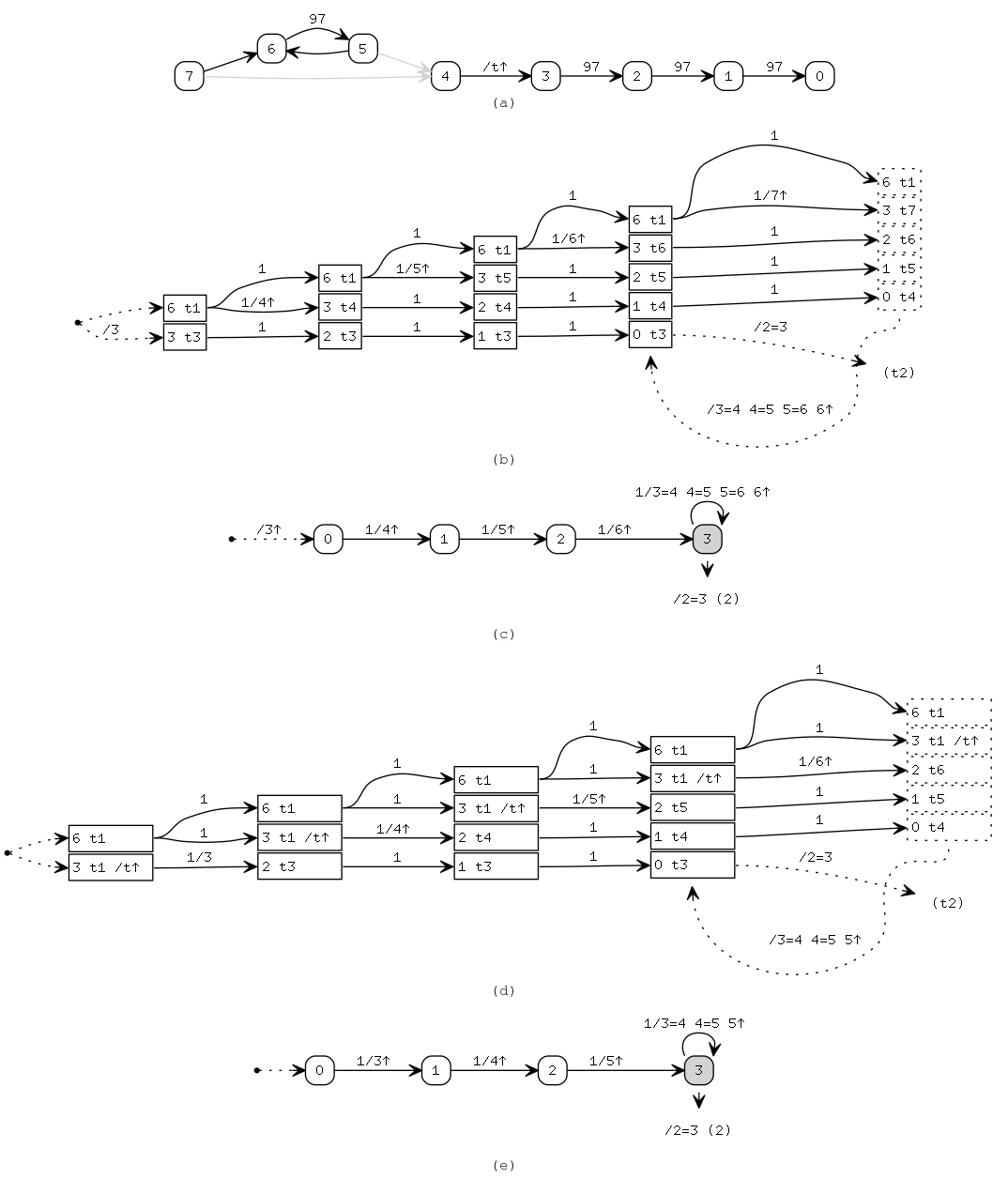}\\*
\textbf{Example 5.} $a^* 1 a^{3}$ .\\*
(a) --- TNFA, (b) --- construction of TDFA(0), (c) --- TDFA(0), (d) --- construction of TDFA(1), (e) --- TDFA(1).\\*
This example demonstrates a pathological case for both types of automata:
nondeterminism degree grows linearly with the number of repetitions.
As a result, for $n$ repetitions both automata contain $O(n)$ states and $O(n)$ copy operations inside of a loop.
TDFA(0) has one more operation than TDFA(1), but for $n \!>\! 2$ this probably makes little difference.
Obviously, for TRE of such kind both methods are impractical.
However, bounded repetition is a problem on its own, even without tags;
relatively small repetition numbers dramatically increase the size of automaton.
If bounded repetition is necessary, more powerful methods should be used:
e.g. automata with \emph{counters} described in \cite{Bec09} (chapter 5.1.12).
\end{Xfig}

\begin{Xfig}
\includegraphics[width=\linewidth]{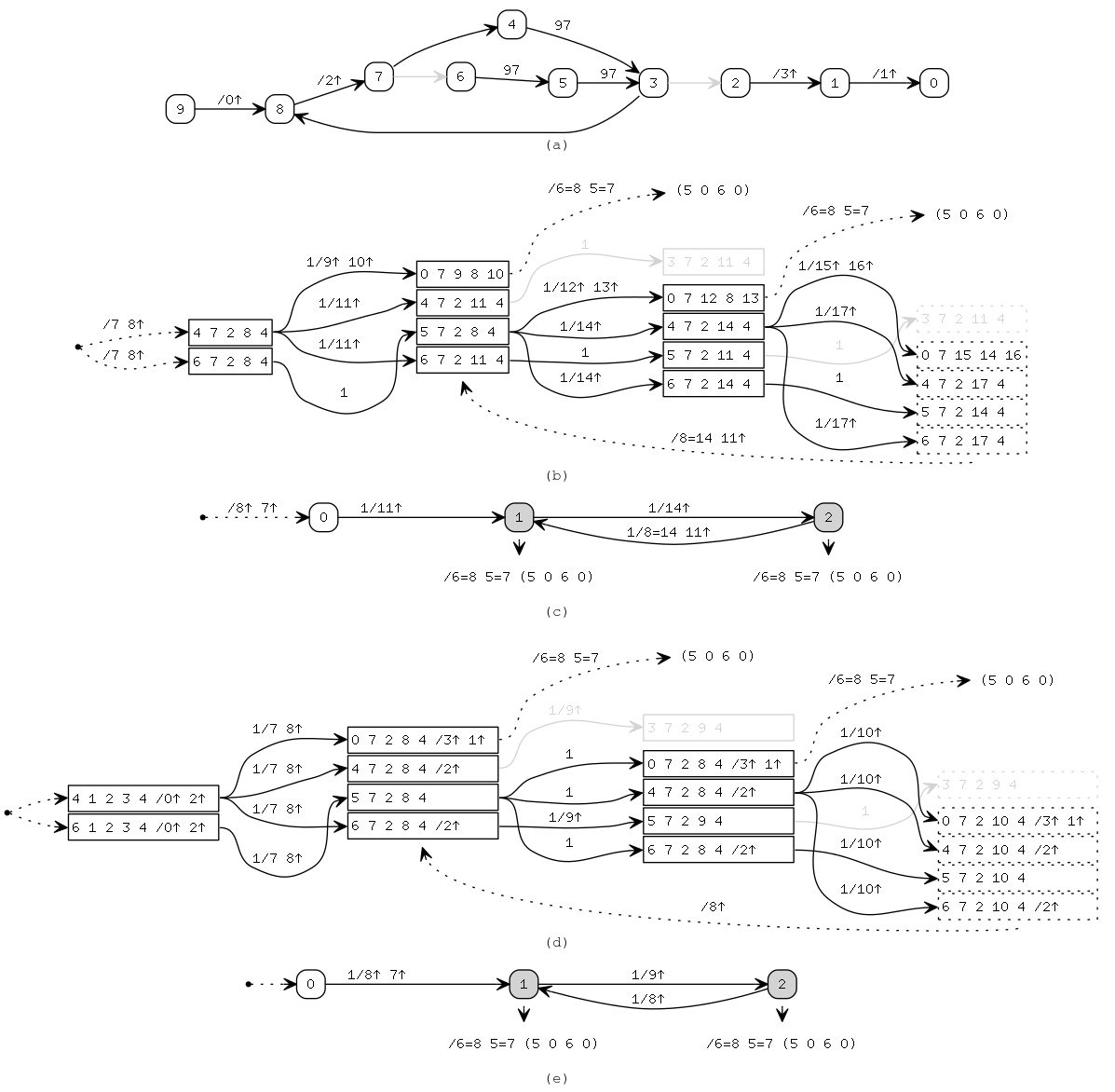}\\*
\textbf{Example 6.} $1 (3 (a | aa) 4)^* 2$, corresponding to POSIX RE \texttt{(a|aa)+}.\\*
(a) --- TNFA, (b) --- construction of TDFA(0), (c) --- TDFA(0), (d) --- construction of TDFA(1), (e) --- TDFA(1).\\*
This example uses POSIX disambiguation.
An early optimization in RE2C rewrites TRE to $1 (3 (a | aa) )^* 4 \, 2$:
orbit tag $4$ is moved out of loop, as we need only its last offset
(disambiguation is based on maximization of tag $3$: as argued in section \ref{section_disambiguation}, checking both tags is redundant).
The resulting automata oscillate between two final states:
submatch result depends on the parity of symbol count in the input string.
Tag $3$ has maximal degree of nondeterminism: $3$ for TDFA(0) and $2$ for TDFA(1).
Tags $2$ and $4$ are deterministic for TDFA(1) and have degree $2$ for TDFA(0).
Tag $1$ is deterministic for both automata.
\end{Xfig}

\begin{multicols}{2}

From these examples we can draw the following conclusions.
First, TDFA(1) is generally better than TDFA(0): delaying register operations allows to get rid of many conflicts.
Second, both kinds of automata are only suitable for RE with modest levels of ambiguity
and low submatch detalisation: TDFA can be applied to full parsing, but other methods would probably outperform them.
However, RE of such form are very common in practice and for them TDFA can be very efficient.



\section{Implementation}\label{section_implementation}

In this section we discuss some practical details that should be taken into account when implementing the above algorithm.
The proposed way of doing things is neither general, nor necessarily the best;
it simply reflects RE2C implementation.

\subsection*{Register reuse}

There are many possible ways to allocate registers during TDFA construction.
One reasonable way (used by Laurikari) is to pick the first register not already used in the given state:
since the number of simultaneously used registers is limited,
it is likely that some of the old ones are not occupied and can be reused.
We use a different strategy: allocate a new register for each distinct operation of each tag on all outgoing transitions from the given state.
It results in a more optimization-friendly automaton
which has a lot of short-lived registers with independent lifetimes.
Consequently, there is less interference between different registers and more registers can be merged.
The resulting program form is similar to \emph{static single assignment} form \cite{SSA},
though not exactly SSA: we cannot use efficient SSA-specific algorithms.
However, SSA construction and deconstruction is rather complex and its usefulness on our (rather simple) programs is not so evident.
\\

It may happen that multiple outgoing transitions from the same state have register operations with identical right-hand sides.
If these operations are induced by the same tag, then one register is allocated for all such transitions.
If, however, operations are induced by different tags, they do not share registers.
But why use different registers, if we know that the same value is written to both of them?
The reason for this is the way we do mapping: if different tags were allowed to share registers,
it would result in a plenty of ``too specialized'' states that do not map to each other.
For example, TDFA for TRE of the form $(1 | \alpha_1) (2 | \alpha_2) \dots (n | \alpha_n)$
would have exponentially many unmappable final states
corresponding to various permutations of default value and current position.

\subsection*{Fallback registers}

So far we have avoided one small, yet important complication.
Suppose that TRE matches two strings, such that one is a proper prefix of the other:
$\alpha_1 \dots \alpha_n$ and $\alpha_1 \dots \alpha_n \beta_1 \dots \beta_m$,
and the difference between them is more than one character: $m \!>\! 1$.
Consider automaton behavior on input string $\alpha_1 \dots \alpha_n \beta_1$:
it will consume all characters up to $\alpha_n$ and arrive at the final state.
Then, however, it will continue matching: since the next character is $\beta_1$, it may be possible to match longer string.
At the next step it will see mismatch and stop.
At that point automaton must backtrack to the latest final state,
restoring input position and all relevant registers that might have been overwritten.
TRE $(a 1 bc)^+$ exhibits this problem for both TDFA(0) and TDFA(1)
(labels 1, 2 and 3 on transitions correspond to symbols \texttt{a}, \texttt{b} and \texttt{c}):

\begin{Xfig}
\includegraphics[width=\linewidth]{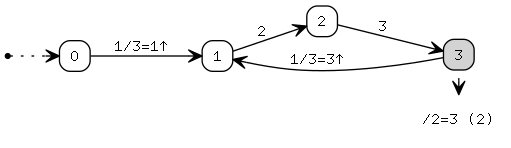}
\captionof{figure}{TDFA(0) for $(a 1 bc)^+$.}
\end{Xfig}

\begin{Xfig}
\includegraphics[width=\linewidth]{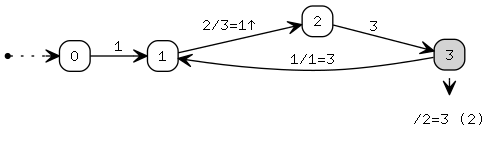}
\captionof{figure}{TDFA(1) for $(a 1 bc)^+$.}
\end{Xfig}

Consider execution of TDFA(0) on input string $abca$: after matching $abc$ in state 3 it will consume $a$ and transition to state 1,
overwriting register 3; then it will fail to match $b$ and backtrack.
Likewise, TDFA(1) will backtrack on input string $abcab$.
Clearly, we must backup register 3 when leaving state 3.
\\

We call registers that need backup \emph{fallback registers}.
Note that not all TRE with overlaps have fallback registers:
it may be that the longer match is unconditional (always matches),
or no registers are overwritten between the two matches,
or the overwritten registers are not used in the final state.
In general, fallback registers can be found by a simple depth-first search from all final states of TDFA.
Each of them needs a \emph{backup register};
all transitions from final state must backup it, and all fallback transitions must restore it.
For the above example the ``repaired'' automata look as follows
(register 3 is renamed to 2, register 1 is backup, fallback transitions are not shown):

\begin{Xfig}
\includegraphics[width=\linewidth]{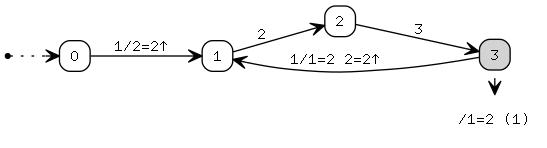}
\captionof{figure}{TDFA(0) for $(a 1 bc)^+$ with backup registers.}
\end{Xfig}

\begin{Xfig}
\includegraphics[width=\linewidth]{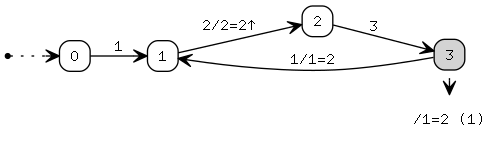}
\captionof{figure}{TDFA(1) for $(a 1 bc)^+$ with backup registers.}
\end{Xfig}

Note that the total number of backup registers cannot exceed the number of tags:
only the latest final state needs to be backup-ed,
and each final TDFA state has only one configuration with final TNFA state,
and this configuration has exactly one register per tag.
As we already allocate distinct final register for each tag,
and this register is not used anywhere else in the program,
we can also use it for backup.

\subsection*{Fixed tags}

It may happen that two tags in TRE are separated by a fixed number of characters:
each offset of one tag is equal to the corresponding offset of the other tag plus some static offset.
In this case we can track only one of the tags; we say that the second tag is \emph{fixed} on the first one.
For example, in TRE $a^* 1 b 2 c^*$ tag 1 is always one character behind of tag 2,
therefore it is fixed on tag 2 with offset -1.
Fixed tags are ubiquitous in TRE that correspond to POSIX RE, because they contain a lot of adjacent tags.
For example, POSIX RE \texttt{(a*)(b*)} is represented with TRE $1 \, 3 \, a^* \, 4 \, 5 \, b^* \, 6 \, 2$,
in which tag 1 is fixed on 3, 4 on 5 and 6 on 2
(additionally, 1 and 3 are always zero and 6, 2 are always equal to the length of matching string).
\\

Fixity relation is transitive, symmetric and reflexive,
and therefore all tags can be partitioned into fixity classes.
For each class we need to track only one representative.
Since fixed tags cannot belong to different alternatives of TRE,
it is possible to find all classes in one traversal of TRE structure
by tracking \emph{distance} to each tag from the nearest non-fixed tag on the same branch of TRE.
Distance is measured as the length of all possible strings that match the part of TRE between two tags:
if this length is variable, distance is infinity and the new tag belongs to a new class.
\\

When optimizing out fixed tags, one should be careful in two respects.
First, negative submatches: if the value of representative is $\varnothing$,
then all fixed tags are also $\varnothing$ and their offsets should be ignored.
Second, fixed tags may be used by disambiguation policy:
in this case they should be kept until disambiguation is finished;
then they can be removed from TDFA with all associated operations.
\\

This optimization is also described in \cite{Lau01}, section 4.3.

\subsection*{Simple tags}

In practice we often need only the last value of some tag:
either because it is not enclosed in repetition and only has one value, or because of POSIX policy, or for any other reason.
We call such tags \emph{simple};
for them determinization algorithm admits a number of simplifications
that result in smaller automata with less register operations.
\\

First, the mapping procedure $map$ from section \ref{section_determinization}
needs not to check bijection between registers if the lookahead history is not empty:
in this case register values will be overwritten on the next step
(for non-simple tags registers would be augmented, not overwritten).
Condition $(v(t), \widetilde{v}(t)) \Xin m(t)$ in the $map$ algorithm on Figure 1
can be replaced with a weaker condition $op(x, t) \!\neq\! \epsilon \vee (v(t), \widetilde{v}(t)) \Xin m(t)$,
which increases the probability of successful mapping.
This optimization applies only to TDFA(1), since lookahead history is always $\epsilon$ for TDFA(0),
so the optimization effectively reduces the gap in the number of states between TDFA(0) and TDFA(1).
\\

Second, operations on simple tags are reduced from normal form $r_1 \Xeq r_2 \cdot b_1 \dots b_n$
to one of the forms $r_1 \Xeq b_n$ (set) and $r_1 \Xeq r_2$ (copy).
It has many positive consequences:
initialization of registers is not necessary;
register values are less versatile and there are less dependencies between registers, therefore more registers can be merged;
operations can be hoisted out of loops.
What is most important, copy operations are cheap for simple tags.

\subsection*{Scalar representation of histories}

The most naive representation of history is a list of offsets;
however, copy operations on lists are very inefficient.
Fortunately, a better representation is possible: as observed by \cite{Kar14}, histories form a \emph{prefix tree}:
each new history is a fork of some old history of the same tag.
Prefix tree can be represented as an array of nodes $(p, o)$,
where $p$ is the index of parent node and $o$ is the offset.
Then each register can hold an index of some leaf node in the prefix tree,
and copy operations are reduced to simple copying of indices.
Append operations are somewhat more complex: they require a new slot (or a couple of slots) in the prefix tree;
however, if array is allocated in large chunks of memory,
then the amortized complexity of each operation is constant.
One inconvenience of this representation is that histories are obtained in reversed form.

\subsection*{Relative vs. absolute values}

If the input is a string in memory, it might be convenient to use \emph{pointers} instead of \emph{offsets}
(especially in C, where all operations with memory are defined in terms of pointers).
However, compared to offsets, pointers have several disadvantages.
First, offsets are usually smaller: often they can be represented with 1-2 bytes, while pointers need 4-8 bytes.
Second, offsets are portable: unlike pointers, they are not tied to a particular environment
and will not loose their meaning if we save submatch results to file or write on a sheet of paper.
Even put aside storage, pointers are sensitive to input buffering:
their values are invalidated on each buffer refill and need special adjustment.
Nevertheless, RE2C uses pointers as default representation of tag values:
this approach is more direct and efficient for simple programs.
RE2C users can redefine default representation to whatever they need.

\subsection*{Optimization pipeline}

Right after TDFA construction and prior to any further optimizations
RE2C performs analysis of unreachable final states
(shadowed by final states that correspond to longer match).
Such states are marked as non-final and all their registers are marked as dead.
\\

After that RE2C performs analysis of fallback registers and adds backup operations as necessary.
\\

Then it applies register optimizations;
they are aimed at reducing the number of registers and copy operations.
This is done by the usual means:
liveness analysis, followed by dead code elimination,
followed by interference analysis and finally register allocation
with biased coalescing of registers bound by copy operations.
The full cycle is run twice (first iteration is enough in most cases,
but subsequent iterations are cheap as they run on an already optimized program and reuse the same infrastructure).
Prior to the first iteration RE2C renames registers so that they occupy consecutive numbers;
this allows to save some space on liveness and interference tables.
\\

Then RE2C performs TDFA minimization:
it is exactly like ordinary DFA minimization, except that
equivalence must take into account register operations:
final states with different finalizers cannot be merged, as well as transitions with different operations.
Thus it is crucial that minimization is applied after register optimizations.
\\

Then RE2C examines TDFA states and, if all outgoing transitions have the same operation,
this operation is hoisted out of transitions into the state itself.
\\

Finally, RE2C converts TDFA to a tunnel automaton \cite{Gro89}
that allows to further reduce TDFA size by merging similar states and deduplicating pieces of code.
\\

Most of these optimizations are basic and some are even primitive, yet put all together and in correct order
they result in a significant reduction of registers, operations and TDFA states
(see the section \ref{section_tests_and_benchmarks} for experimental results).

\section{Tests and benchmarks}\label{section_tests_and_benchmarks}

\subsection*{Correctness}

Correctness testing of RE2C was done in several different ways.
First, about a hundred of hand-written tests were added to the main RE2C test suite.
These tests include examples of useful real-world programs
and checks for various optimizations, errors and special cases.
\\

Second, RE2C implementation of POSIX captures was verified on the canonical POSIX test suite composed by Glenn Fowler \cite{Fow03}.
I used the augmented version provided by Kuklewicz \cite{Kuk09} and excluded a few tests that check POSIX-specific extensions
which are not supported by RE2C (e.g. start and end anchors \texttt{\^} and \texttt{\$}) ---
the excluded tests do not contain any special cases of submatch extraction.
\\

Third, and probably most important, I used the \emph{fuzzer} contributed by Sergei Trofimovich
(available as a part of RE2C source code)
and based on the Haskell QuickCheck library \cite{CH11}.
Fuzzer generates random RE with the given \emph{constrains}
and verifies that each generated RE satisfies certain \emph{properties}.
By redefining the set of constraints one can control the size and the form of RE:
for example, tweak the probability of different operations or change the basic character set.
One can tune fuzzer to emit RE with heavy use of some particular feature,
which is often useful when testing various implementation aspects.
Properties, on the other hand, control the set of tests and checks that are applied to each RE:
by redefining properties it is possible to chase all sorts of bugs.
\\

While RE were generated at random, each particular RE was tested extensively
on the set of input strings generated with RE2C \texttt{--skeleton} option.
This option enables RE2C self-validation mode:
instead of embedding the generated lexer in used-defined interface code,
RE2C embeds it in a self-contained template program called \emph{skeleton}.
Additionally, RE2C generates two input files: one with strings derived from the regular grammar
and one with compressed match results that are used to verify skeleton behavior on all inputs.
Input strings are generated so that they cover all TDFA transitions and many TDFA paths
(including paths that cause match failure).
Data generation happens right after TDFA construction and prior to any optimizations,
but the lexer itself is fully optimized (it is the same lexer that would be generated in normal mode).
Thus skeleton programs are capable of revealing any errors in optimization and code generation.
\\

Combining skeleton with fuzzer yields a powerful and generic testing method.
I used it to verify the following properties:

\begin{itemize}
    \setlength{\parskip}{0.5em}

    \item Correctness of RE2C optimizations:
        fuzzer found tens of bugs in the early implementation of tags in RE2C,
        including some quite involved and rare bugs that occurred on later stages of optimization
        and would be hard to find otherwise.

    \item Coherence of TDFA(0) and TDFA(1):
        the two automata result in different programs which must yield identical results.
        I ran TDFA(0) programs on skeleton inputs generated for TDFA(1) programs and vice versa;
        it helped to reveal model-specific bugs.

    \item Coherence of RE2C and Regex-TDFA (Haskell RE library written by Kuklewicz that supports POSIX submatch semantics \cite{Regex-TDFA}).
        I ran Regex-TDFA on skeleton input strings generated by RE2C and compared match results with those of the skeleton program.
        Aside from a couple of minor discrepancies (such as newline handling and anchors)
        I found two bugs in submatch extraction in Regex-TDFA.
        Both bugs were found multiple times on slightly different RE and inputs,
        and both are relatively rare (the faulty RE occurred approximately once in 50 000 tests
        and it only failed on some specific input strings).
        On the bulk of inputs RE2C and Regex-TDFA are coherent.

        First bug can be triggered by RE \texttt{(((a*)|b)|b)+} and input string \texttt{ab}:
        Regex-TDFA returns incorrect submatch result for second capturing group \texttt{((a*)|b)}
        (no match instead of \texttt{b} at offset 1).
        Some alternative variants that also fail: \texttt{(((a*)|b)|b)\{1,2\}}, \texttt{((b|(a*))|b)+}.

        Second bug can be triggered by RE \texttt{((a?)(())*|a)+} and input string \texttt{aa}.
        Incorrect result is for second group \texttt{(a?)} (no match instead of \texttt{a} at offset 1),
        third group \texttt{(())} and fourth group \texttt{()} (no match instead of empty match at offset 2).
        Alternative variant that also fails: \texttt{((a?()?)|a)+}.

        Tested against Regex-TDFA-1.2.2.

    \item Numerous assumptions and hypotheses that arose during this work:
        fuzzer is a most helpful tool to verify or disprove one's intuition.
    \\
\end{itemize}

I did not compare RE2C against other libraries, such as \cite{TRE} or \cite{RE2},
as none of these libraries support POSIX submatch semantics:
TRE has known bugs \cite{LTU},
and RE2 author explicitly states that POSIX submatch semantics is not supported \cite{Cox17}.

\subsection*{Benchmarks}

Benchmarks are aimed at comparison of TDFA(0) and TDFA(1);
comparison of RE2C and other lexer generators is beyond the scope of this paper (see \cite{BC93}).
As we have already seen on numerous examples in section \ref{section_determinization},
TDFA(1) has every reason to result in faster code;
however, only a real-world program can show if there is any perceivable difference in practice.
I used two canonical use cases for submatch extraction in RE: URI parser and HTTP parser.
Both examples are used in literature \cite{BT10} \cite{GHRST16},
as they are simple enough to admit regular grammar,
but at the same time both grammars have non-trivial structure composed of multiple components of varying length and form \cite{RFC-3986} \cite{RFC-7230}.
Each example has two implementations: RFC-compliant and simplified (both forms may be useful in practice).
The input to each parser is a 1G file of randomly generated URIs or HTTP messages; it is buffered in 4K chunks.
Programs are written so that they spend most of the time on parsing,
so that benchmarks measure the efficiency of parsing, not the accompanying code or the operating system.
For each of the four parsers there is a corresponding DFA-based recognizer:
it sets a baseline for expectations of how fast and small the lexer can be and what is the real overhead on submatch extraction.
Benchmarks are written in C-90 and compiled with \cite{RE2C} version 1.0
and four different C compilers:
\cite{GCC} version 7.1.10,
\cite{Clang} version 4.0.1,
\cite{TCC} version 0.9.26
and \cite{PCC} version 1.1.0
with optimization level \texttt{-O2} (though some compilers probably ignore it).
RE2C was run in three different settings:
default mode, with \texttt{-b} option (generate bit masks and nested \texttt{if}-s instead of plain \texttt{switch}-es),
and with \texttt{--no-optimize-tags} option (suppress optimizations of tag variables described in section \ref{section_implementation}).
All benchmarks were run on 64-bit Intel Core i3 machine with 3G RAM and 32K L1d, 32K L1i, 256K L2 and 3072K L3 caches;
each result is the average of 4 subsequent runs after a proper warm-up.
Benchmark results are summarized in tables 1 --- 4
and visualized on subsequent plots.
\\

Benchmarks are available as part of RE2C-1.0 distribution
in subdirectory \texttt{re2c/benchmarks}.

\end{multicols}

\begin{Xtab}\label{table1}
\begin{center}
    \begin{tabular}{|c|ccccccccccc|}
    \hline
    & registers & states & code size (K) & \multicolumn{4}{c}{stripped binary size (K)} & \multicolumn{4}{c|}{run time (s)} \\
    & & &
        & gcc & clang & tcc & pcc
        & gcc & clang & tcc & pcc \\
    \hline \hline
    \multicolumn{12}{|c|}{re2c} \\
    \hline
    TDFA(0) & 45 & 452 & 250 & 63 & 135 & 339 & 247 & 12.86 & 10.27 & 99.09 & 55.83 \\
    TDFA(1) & 42 & 457 & 183 & 55 & 139 & 213 & 151 &  6.43 &  5.59 & 67.00 & 27.93 \\
    DFA     & -- & 414 & 135 & 35 & 111 & 145 &  91 &  4.96 &  4.46 & 62.04 & 23.67 \\
    \hline \hline
    \multicolumn{12}{|c|}{re2c -b} \\
    \hline
    TDFA(0) & 45 & 452 & 295 & 63 & 59 & 352 & 267 & 11.95 & 10.30 & 65.47 & 36.95 \\
    TDFA(1) & 42 & 457 & 171 & 55 & 51 & 144 & 111 &  6.01 &  5.40 & 15.94 & 10.53 \\
    DFA     & -- & 414 & 123 & 35 & 39 &  75 &  51 &  4.71 &  4.76 & 10.88 &  5.61 \\
    \hline \hline
    \multicolumn{12}{|c|}{re2c --no-optimize-tags} \\
    \hline
    TDFA(0) & 2054 & 625 & 816 & 275 & 267 & 1107 & 839 & 14.11 & 13.25 & 105.58 & 59.60 \\
    TDFA(1) &  149 & 462 & 200 &  63 & 147 &  233 & 167 &  6.47 &  5.90 &  68.43 & 29.09 \\
    \hline
    \end{tabular}
    \captionof{table}{RFC-7230 compliant HTTP parser.\\
    Total 39 tags: 34 simple and 5 with history.
    Nondeterminism for TDFA(0): 23 tags with degree 2, 12 tags with degree 3 and 1 tag with degree 4.
    Nondeterminism for TDFA(1): 18 tags with degree 2, 2 tags with degree 3.}
\end{center}
\end{Xtab}

\begin{Xtab}\label{table2}
\begin{center}
    \begin{tabular}{|c|ccccccccccc|}
    \hline
    & registers & states & code size (K) & \multicolumn{4}{c}{stripped binary size (K)} & \multicolumn{4}{c|}{run time (s)} \\
    & & &
        & gcc & clang & tcc & pcc
        & gcc & clang & tcc & pcc \\
    \hline \hline
    \multicolumn{12}{|c|}{re2c} \\
    \hline
    TDFA(0) & 18 & 70 & 32 & 15 & 31 & 41 & 31 & 7.66 & 5.47 & 71.60 & 33.90 \\
    TDFA(1) & 16 & 73 & 33 & 15 & 35 & 41 & 31 & 5.30 & 3.83 & 63.30 & 26.74 \\
    DFA     & -- & 69 & 25 & 15 & 31 & 31 & 23 & 4.90 & 3.34 & 62.00 & 23.59 \\
    \hline \hline
    \multicolumn{12}{|c|}{re2c -b} \\
    \hline
    TDFA(0) & 18 & 70 & 31 & 15 & 19 & 31 & 31 & 7.12 & 7.30 & 31.81 & 17.44 \\
    TDFA(1) & 16 & 73 & 29 & 15 & 19 & 29 & 27 & 5.24 & 4.43 & 13.50 &  8.84 \\
    DFA     & -- & 69 & 19 & 11 & 15 & 15 & 15 & 4.64 & 3.94 & 11.00 &  5.77 \\
    \hline \hline
    \multicolumn{12}{|c|}{re2c --no-optimize-tags} \\
    \hline
    TDFA(0) & 72 & 106 & 57 & 23 & 55 & 73 & 55 & 8.61 & 6.77 & 72.96 & 34.63 \\
    TDFA(1) & 44 &  82 & 39 & 19 & 43 & 49 & 39 & 6.00 & 5.39 & 63.79 & 27.37 \\
    \hline
    \end{tabular}
    \captionof{table}{Simplified HTTP parser.\\
    Total 15 tags: 12 simple and 3 with history.
    Nondeterminism for TDFA(0): 8 tags with degree 2.
    Nondeterminism for TDFA(1): 3 tags with degree 2.}
\end{center}
\end{Xtab}

\begin{Xtab}\label{table3}
\begin{center}
    \begin{tabular}{|c|ccccccccccc|}
    \hline
    & registers & states & code size (K) & \multicolumn{4}{c}{stripped binary size (K)} & \multicolumn{4}{c|}{run time (s)} \\
    & & &
        & gcc & clang & tcc & pcc
        & gcc & clang & tcc & pcc \\
    \hline \hline
    \multicolumn{12}{|c|}{re2c} \\
    \hline
    TDFA(0) & 23 & 252 & 152 & 39 & 75 & 203 & 155 & 10.01 & 6.01 & 111.76 & 73.75 \\
    TDFA(1) & 20 & 256 & 115 & 35 & 75 & 138 & 103 &  6.78 & 3.23 & 104.36 & 51.00 \\
    DFA     & -- & 198 &  67 & 23 & 55 &  73 &  55 &  7.06 & 3.19 &  97.87 & 51.37 \\
    \hline \hline
    \multicolumn{12}{|c|}{re2c -b} \\
    \hline
    TDFA(0) & 23 & 252 & 165 & 39 & 35 & 181 & 151 & 8.36 & 8.58 & 39.51 & 31.81 \\
    TDFA(1) & 20 & 256 & 127 & 55 & 31 & 130 & 107 & 5.21 & 4.81 & 12.02 & 10.01 \\
    DFA     & -- & 198 &  60 & 19 & 23 &  39 &  35 & 4.04 & 4.06 &  9.13 &  8.17 \\
    \hline \hline
    \multicolumn{12}{|c|}{re2c --no-optimize-tags} \\
    \hline
    TDFA(0) & 611 & 280 & 426 & 127 & 151 & 536 & 463 & 10.39 & 7.51 & 127.35 & 75.23 \\
    TDFA(1) &  64 & 256 & 131 &  43 &  87 & 156 & 123 &  6.74 & 3.54 & 103.91 & 51.08 \\
    \hline
    \end{tabular}
    \captionof{table}{RFC-3986 compliant URI parser.\\
    Total 20 tags (all simple).
    Nondeterminism for TDFA(0): 15 tags with degree 2 and 4 tags with degree 3.
    Nondeterminism for TDFA(1): 10 tags with degree 2.}
\end{center}
\end{Xtab}

\begin{Xtab}\label{table4}
\begin{center}
    \begin{tabular}{|c|ccccccccccc|}
    \hline
    & registers & states & code size (K) & \multicolumn{4}{c}{stripped binary size (K)} & \multicolumn{4}{c|}{run time (s)} \\
    & & &
        & gcc & clang & tcc & pcc
        & gcc & clang & tcc & pcc \\
    \hline \hline
    \multicolumn{12}{|c|}{re2c} \\
    \hline
    TDFA(0) & 16 & 26 & 17 & 11 & 19 & 23 & 19 & 8.34 & 3.55 & 102.72 & 59.84 \\
    TDFA(1) & 13 & 28 & 19 & 11 & 19 & 25 & 23 & 6.04 & 3.12 & 100.28 & 47.85 \\
    DFA     & -- & 22 & 10 & 11 & 15 & 14 & 15 & 5.89 & 2.66 &  97.95 & 47.01 \\
    \hline \hline
    \multicolumn{12}{|c|}{re2c -b} \\
    \hline
    TDFA(0) & 16 & 26 & 20 & 11 & 11 & 22 & 23 & 7.14 & 6.67 & 23.19 & 18.73 \\
    TDFA(1) & 13 & 28 & 17 & 11 & 11 & 19 & 19 & 4.02 & 3.08 &  8.56 &  6.90 \\
    DFA     & -- & 22 &  7 & 11 & 11 &  8 & 11 & 3.90 & 2.52 &  8.00 &  4.40 \\
    \hline \hline
    \multicolumn{12}{|c|}{re2c --no-optimize-tags} \\
    \hline
    TDFA(0) & 79 & 29 & 33 & 19 & 23 & 43 & 39 & 7.43 & 4.05 & 105.06 & 61.74 \\
    TDFA(1) & 40 & 31 & 28 & 15 & 23 & 36 & 31 & 6.27 & 3.32 & 101.79 & 48.15 \\
    \hline
    \end{tabular}
    \captionof{table}{Simplified URI parser.\\
    Total 14 tags (all simple).
    Nondeterminism for TDFA(0): 8 tags with degree 2 and 5 tags with degree 3.
    Nondeterminism for TDFA(1): 7 tags with degree 2.}
\end{center}
\end{Xtab}

\begin{Xfig}
\includegraphics[width=\linewidth]{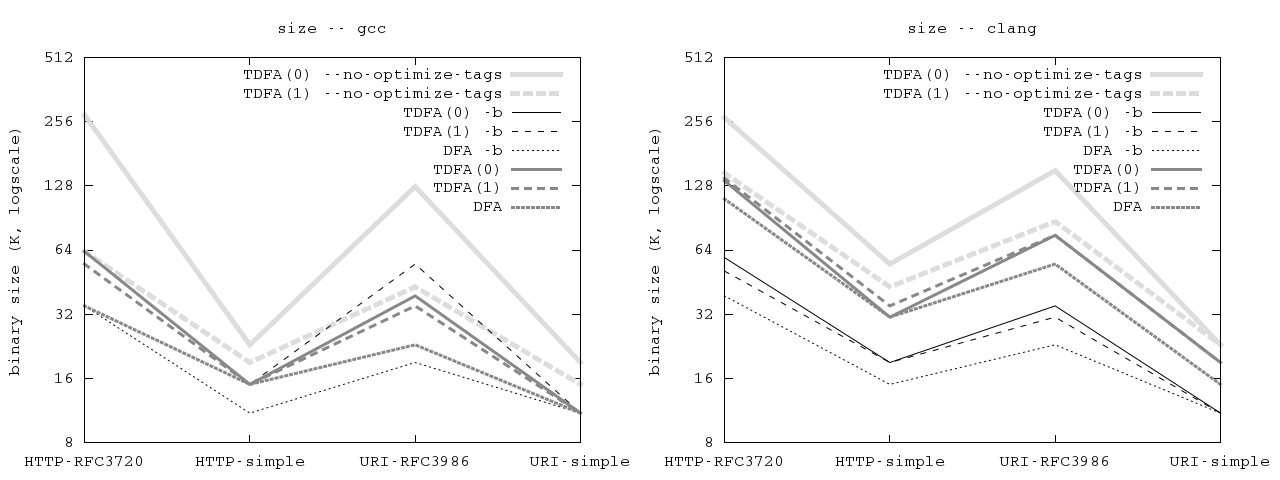}
\includegraphics[width=\linewidth]{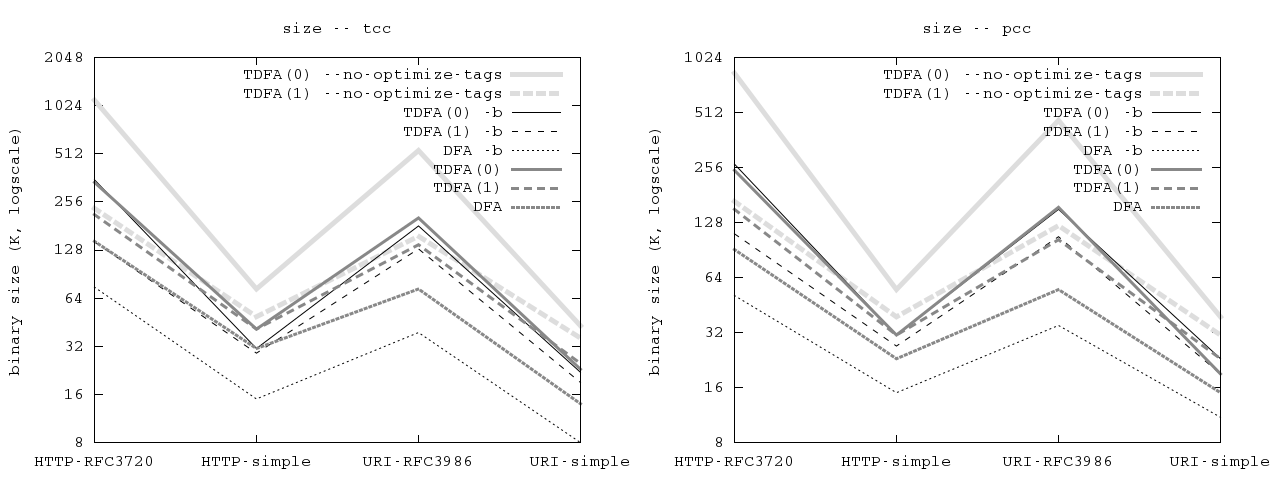}
\captionof{figure}{Binary size for GCC, Clang, TCC and PCC.}
\end{Xfig}

\begin{Xfig}
\includegraphics[width=\linewidth]{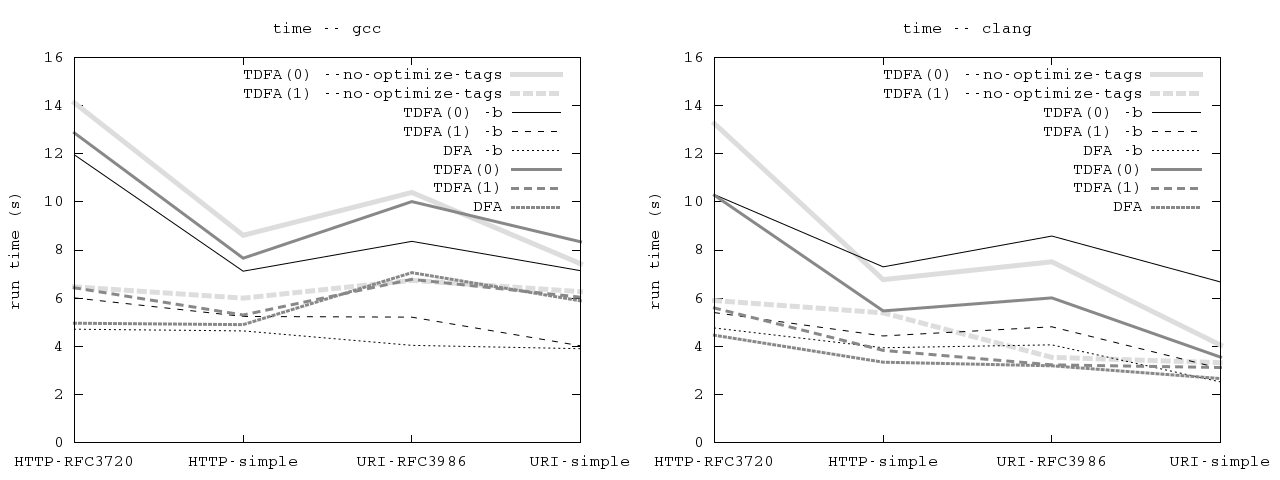}
\includegraphics[width=\linewidth]{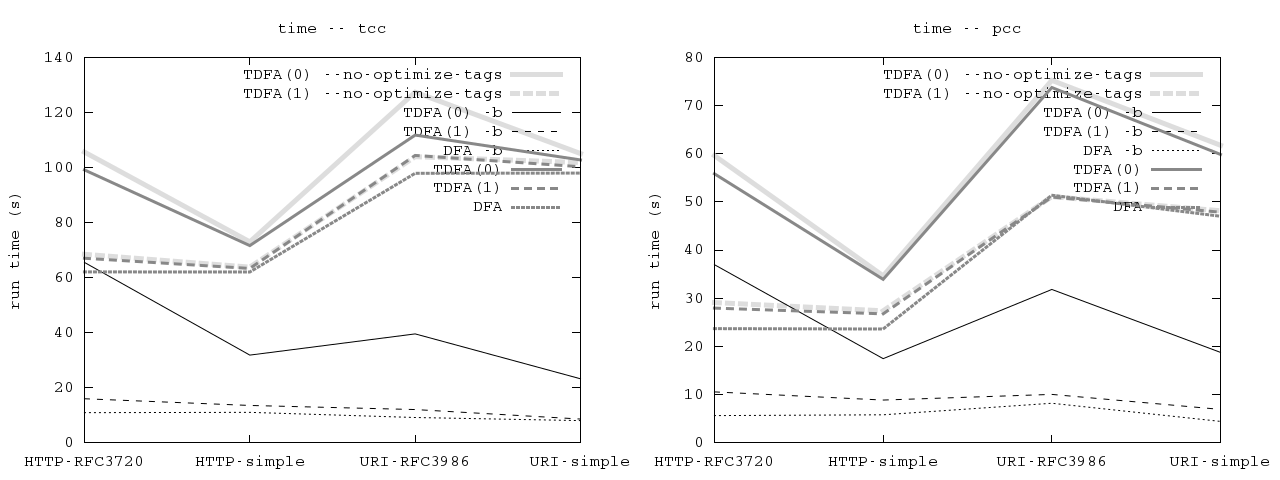}
\captionof{figure}{Run time for GCC, Clang, TCC and PCC.}
\end{Xfig}

\begin{multicols}{2}

Benchmark results show the following:

\begin{itemize}
    \setlength{\parskip}{0.5em}

    \item Speed and size of the generated code vary between different compilers:
        as expected, TCC and PCC generate slower and larger code than GCC and Clang (though PCC performs notably better);
        but even GCC and Clang, which are both known for their optimizations, generate very different code:
        GCC binaries are often 2x smaller, while the corresponding Clang-generated code runs up to 2x faster.

    \item RE2C code-generation option \texttt{-b} has significant impact on the resulting code:
        it results in up to 5x speedup for TCC, 2x speedup for PCC and about 2x reduction of binary size for Clang at the cost of about 1.5x slowdown;
        of all compilers only GCC seems to be unaffected by this option.

    \item Regardless of different compilers and options, TDFA(1) is consistently more efficient than TDFA(0):
        the resulting code is about 1.5 - 2x faster and generally smaller,
        especially on large programs and in the presence of tags with history.

    \item TDFA(1) incurs modest overhead on submatch extraction compared to DFA-based recognition;
        in particular, the gap between DFA and TDFA(0) is smaller than the gap between TDFA(0) and TDFA(1).

    \item Nondeterminism levels are not so high in the example programs.

    \item RE2C optimizations of tag variables reduce binary size, even with optimizing C compilers.

    \item RE2C optimizations of tag variables have less effect on execution time: usually they reduce it, but not by much.
    \\
\end{itemize}

\section{Conclusions}\label{section_conclusions}

TDFA(1) is a practical method for submatch extraction in lexer generators that optimize for speed of the generated code.
It incurs a modest overhead compared to simple recognition,
and the overhead depends on detalization of submatch
(in many cases it is proportional to the number of tags).
One exception is the case of ambiguous submatch in the presence of bounded repetition:
it causes high degree of nondeterminism for the corresponding tags
and renders the method impractical compared to hand-written code.
\\ \\
TDFA(1) method is considerably more efficient than TDFA(0) method, both theoretically and practically.
Experimental results show that TDFA(1) achieves 1.5x -- 2x speedup compared to TDFA(0)
and in most cases it results in smaller binary size.
\\ \\
TDFA method is capable of extracting repeated submatches,
and therefore it is applicable to full parsing.
Efficiency of the generated parsers depends on the data structures used to hold and manipulate repeated submatch values
(an efficient implementation is possible).
\\ \\
TDFA can be used in combination with various disambiguation policies;
in particular, leftmost greedy and POSIX policies.

\section{Future work}\label{section_future_work}

The most interesting subject that needs further exploration and experiments
is the comparison of TDFA (described in this paper) and DSST (described in \cite{Gra15} and \cite{GHRST16})
on practical problems of submatch extraction.
Both models are aimed at generating fast parsers,
and both depend heavily on the efficiency of particular implementation.
For instance, DSST is applied to full parsing, which suggests that it has some overhead on submatch extraction compared to TDFA;
however, optimizations of the resulting program may reduce the overhead, as shown in \cite{Gra15}.
On the other hand, TDFA allows copy operations on registers, contrary to DSST;
but in practice copy operations are cheap if the registers hold scalar values, as shown in section \ref{section_implementation}.
The author's expectation is that on RE of modest size and submatch complexity
optimized implementations of TDFA and DSST should result in very similar code.
The construction of DSST given in \cite{Gra15} works only for leftmost greedy disambiguation;
it might be interesting to construct DSST with POSIX disambiguation.
\\ \\
Extending TDFA lookahead to more than one symbol (in other words, extending TDFA to \emph{multi-stride} automata described in \cite{Bec09})
is an interesting theoretical experiment, but probably not very useful in practice.
As in the case of LR($k$) methods for $k > 1$, TDFA($k$) would pobably be much larger and yet insufficiently expressive to resolve all conflicts.
\\ \\
A more practical subject is combining TDFA and the \emph{counting automata} described in \cite{Bec09}:
it would solve the problem of tag nondeterminism in the presence of bounded repetition.
\\ \\
It would be interesting to implement more involved analysis and optimizations in RE2C,
as it has stronger guarantees and deeper knowledge of the program than the C compiler.

\section*{Acknowledgments}

This study would not be possible without the help of Sergei Trofimovich.
His relentless work on open source projects
and his ability to track down and fix the hardest bugs are my highest ideal of a programmer.
If it were not for him, I would not even know about RE2C.
\\

All that I understand in mathematics I owe to my parents Vladimir Fokanov and Elina Fokanova,
my school teacher Tatyana Leonidovna Ilyushenko
and the Belarusian State University [BSU].
\\

And many thanks to all the good people who cheered me up during this work. :)

\end{multicols}
\pagebreak

\nocite{*}

\newcommand{\etalchar}[1]{$^{#1}$}

\end{document}